\documentclass[10pt,letterpaper]{article}

\usepackage{hyperref}
\usepackage{nameref}
\usepackage[top=0.85in,left=2.75in,footskip=0.75in]{geometry}

\usepackage{amsmath}
\usepackage{amssymb}

\usepackage{changepage}
\usepackage{placeins}

\usepackage[utf8x]{inputenc}

\usepackage{textcomp,marvosym}
\usepackage{epsfig}

\usepackage{cite}

\usepackage[usenames]{color}
\usepackage{ulem}
\usepackage{microtype}
\DisableLigatures[f]{encoding = *, family = * }

\usepackage[table]{xcolor}

\usepackage{array}
\usepackage{graphicx}
\newcolumntype{+}{!{\vrule width 2pt}}

\newlength\savedwidth

\usepackage[aboveskip=1pt,labelfont=bf,labelsep=period,justification=raggedright,singlelinecheck=off]{caption}

\makeatletter
\renewcommand{\@biblabel}[1]{\quad#1.}
\makeatother

\date{}

\usepackage{lastpage,fancyhdr,graphicx}
\usepackage{epstopdf}
\fancyheadoffset[L]{2.25in}
\fancyfootoffset[L]{2.25in}
\definecolor{afx}{rgb}{0.4, 0.17, 0.89}

\RequirePackage{float}
\RequirePackage[abs]{overpic}
\usepackage{color}

\begin{document}
\vspace*{0.2in}

\begin{flushleft}	
	
{\Large\textbf{Collective Information Processing}}
\vskip 0.14cm	
{\Large\textbf{in Human Phase Separation}}
\newline

{Bertrand Jayles$^{1,2}$, Ram\'on Escobedo$^2$, Roberto Pasqua$^3$,
Christophe Zanon$^3$,  Adrien Blanchet$^{4,5}$, Matthieu Roy$^3$, Gilles Tredan$^3$, 
Guy Theraulaz$^{2,5}$, 	Cl\'ement Sire$^{1,*}$}
\bigskip
\date{}

{\small{$^1$Laboratoire de Physique Th\'eorique, CNRS, Universit\'e de Toulouse -- Paul Sabatier, France\\
		$^2$Centre de Recherches sur la Cognition Animale, Centre de Biologie Int\'egrative,  
		 CNRS, Universit\'e de Toulouse  -- Paul Sabatier, France\\
		$^3$Laboratoire d'Analyse et d'Architecture des Syst\`emes, CNRS, Universit\'e de Toulouse, France\\
		$^4$Toulouse School of Economics, CNRS, Universit\'e de Toulouse  -- Capitole,  France\\
		$^5$Institute for Advanced Study in Toulouse, France}}
\bigskip

{*\,Clement.Sire@irsamc.ups-tlse.fr}

\end{flushleft}

%\linenumbers
\section*{Abstract}
Social media filters combined with recommender systems can lead to the emergence of filter bubbles and polarized groups. In addition, segregation processes of human groups in certain social contexts have been shown to share some similarities with phase separation phenomena in physics.  Here, we study the impact of information filtering on collective segregation behavior. We report a series of experiments where groups of 22 subjects have to perform a collective segregation task  that mimics the tendency of individuals to bond with other similar individuals. More precisely, the participants are each assigned a color (red or blue) unknown to them, and have to regroup with other subjects sharing the same color. To assist them, they are equipped with an artificial sensory device capable of detecting the majority color in their ``environment'' (defined as their $k$ nearest neighbors, unbeknownst to them), for which we control the perception range, $k=1,3,5,7,9,11,13$. We study the separation dynamics (emergence of unicolor groups) and the properties of the final state, and show that the value of $k$ controls the quality of the segregation, although the subjects are totally unaware of the precise definition of the ``environment''. We also find that there is a perception range $k=7$ above which the ability of the group to segregate does not improve. We introduce a model that precisely describes the random motion of a group of pedestrians in a confined space, and which faithfully reproduces and allows to interpret the results of the segregation experiments. Finally, we discuss the strong and precise analogy between our experiment and the phase separation of two immiscible materials at very low temperature.

\vskip 0.5cm
\noindent \textbf{Keywords:} {collective human  behavior $|$ collective information processing $|$ collective motion $|$ phase separation $|$ computational modeling}

\newpage

\section*{Author Summary}
In our digital societies, one of the negative consequences of information filtering is the formation of polarized groups. We have investigated the impact of the amount of information gathered locally by individuals on the collective segregation dynamics in human groups. We performed experiments in which groups of red or blue subjects had to separate into unicolor groups. The individuals were equipped with an artificial sensory device filtering the amount of information on their environment to which they had access, and delivering a single-bit acoustic signal in return. We show that the nature of this signal controls the dynamics and the quality of the group segregation, which does not improve beyond the acquisition of a critical amount of information.

\section*{Introduction}
	
The development of the Internet, social media, and mobile technologies has  changed the way humans communicate, collaborate, consume, and create information \cite{bargh_internet_2004,schoder_social_2013,castells_end_2010,happer_role_2013}. People live in a digital world in which they are more connected  than ever, in an era where knowledge has never been so accessible, and where available viewpoints, perspectives, ideas, and opinions are more diverse than ever \cite{negroponte_being_1995}. Moreover, through the use of recommender systems,  information technologies help customers to make their choices using the experience and feedback of other users \cite{ricci_recommender_2011}.

Nevertheless, these technologies have also brought some negative side effects for individual users and society as a whole. Because of the overabundance of information and the limited time and attention people possess, online media providers exploit automatic filtering algorithms that provide different and tailored information to individual users \cite{mobasher_automatic_2000}. While useful, these social media filters have contributed to the amplification of homophily and the tendency of individuals to selectively expose themselves to the opinion of like-minded people, thus reinforcing their existing beliefs and contributing to the emergence of filter bubbles or echo chambers \cite{pariser_filter_2011,bakshy_exposure_2015,nikolov_measuring_2015,bozdag2015}. This raises concerns that information technologies can eventually lead to increasingly polarized and fragmented societies \cite{sunstein_Republic_2018}.
Interestingly, the formation of polarized groups as a consequence of information filtering displays many similarities with the phase separation of two immiscible materials in physical and biological systems \cite{khabibullaev_phase_2013,hyman_liquid_2014,castellano_statistical_2009,majsire}. 
Yet, an important issue that has received little attention so far
is the impact that the amount of information perceived and used at the individual level has on the segregation dynamics and the resulting fragmentation of groups. 
Understanding this impact could help to shed light on the conditions under which one could eventually control the formation of polarized groups. 

To investigate this question in a well controlled setup, we have performed a series of experiments with human groups that mimic the tendency of individuals to bond with other similar individuals. 
The computer controlling the experiment  first assigns a color --  red or blue -- to each individual in a group,  unknown to him/her and others, but fixed in a given experimental run. 
We use an electronic sensory device to control the amount of information about the environment delivered to the individuals without them being aware of the nature of this control. Each participant wears location tags allowing the computer to collect in real-time the information about the positions and colors of their $k$ nearest neighbors, the value of $k$ being controlled by and only known to the experimenter. The sensory device determines the color of the majority of these $k$ neighbors and the tags deliver an acoustic signal (a ``beep'') to the subject if this dominant color is different from her/his own color.  An experimental run ends when all sensory devices have stopped beeping, which corresponds to the formation of groups composed of individuals of the same color, resulting in a partial or complete segregation. The simple nature of the signal -- the one-bit beep -- allows the individuals to be free from any cognitive overload linked to the acquisition and processing of a growing volume of information on their environment.

Whatever the experimental conditions (the value of $k$), the signal of the sensory device is perceived by the subjects in the same way, despite being based on the acquisition of a varying  amount of information. Yet, we will show that different values of $k$ can lead to diverse forms of group organization characterized by very different levels of segregation.   Moreover, we have built a model which precisely describes the random motion of a group of pedestrians in a confined space and which faithfully reproduces and explains the different features of our experiment. Our analysis also points to an intriguing and precise analogy between these experiments and the phase separation of two immiscible materials at very low temperature.

\section*{Results}

\paragraph*{Fig~1.}
{\label{fig:1}\small\textbf{Illustration of the different stages of an experimental run.} (\textbf{a})~Random walk phase for typically 45 seconds (top panel), phase separation process (beeping phase; middle panel), and final silent state where the subjects do not beep and do not move anymore (bottom panel), exemplified for an experiment with $k = 5$.
	(\textbf{b})~Decay of the fraction of beeping subjects (purple lines) and of the number of 3-groups (black lines) averaged over all experimental runs for $k=1$, 3, 5, 7, 9, and 11, compared to the results of the model (red smooth lines).}\vskip 0.4cm

\subsection*{Experiments}
Prior to each experiment, the participants were told that they had each been  allocated a color (red or blue), unknown to them (11 subjects of each color). 
They were also informed that after an initial period of typically 45\,s during  which they were asked to walk randomly and at their normal pace in the circular arena of radius $R=3.56$\,m (\nameref{fig:1}a top panel), the tag clipped on their left shoulder would start beeping \textit{if the subjects in their environment were not the same color as them} (verbatim; \nameref{fig:1}a middle panel, and 
\nameref{setup} and \nameref{beepingcases}). 
Ultimately, they were told that a run of the experiment would stop only after all subjects had stopped beeping (\nameref{fig:1}a bottom panel), after which another run of the experiment would start again (see \nameref{Table1}).

\paragraph*{Fig~2.}
{\small\textbf{Segregation experiment and tracking system}. An experimental run starts with the random
	color assignment to the 22~participants (11~blue and 11~red) and the random walking
	phase (a), where participants walk in a circular arena delimited with a blue tape on the floor.
	In the segregation phase (b, c, e), the tag clipped on each participant left shoulder emits
	a beep if the color of the majority of his/her $k$ nearest neighbors ($k = 1$, 3, 5, 7, 9, 11, and 13) is different from 	his/her own. The positions of the participants are  recorded in real time by the Ubisense tracking system (c, d) based on the triangulation of sensors (d) located around the	arena.}
\label{setup}\vskip 0.3cm

\paragraph*{Fig~3.}
{\small\textbf{Illustration of the beeping state}:
	the red focal individual is marked by a circle, and is connected to its $k$ nearest neighbors by arrows. 
	The tags on his/her left shoulder beeps when the majority of his/her $k$ nearest neighbors is of a different color from him/her (for $k = 3$ and 7 above, as illustrated by the blue wave-like symbol). 
	Grey neighbors are beyond the interaction range, and are thus not considered in the ‘‘environment’’ of the focal individual (i.e., in the computation of the beep).}
\label{beepingcases}\vskip 0.4cm

Despite the fact that our purposely vague phrasing did not explicit the nature of the ``environment'', no subject ever asked further details about it. Yet, the precise definition of this environment was central to our experiment. For a given odd $k=1,3,5,9,11,13$, the left tag of a subject would beep if the majority of his/her $k$ nearest neighbors were \textit{not} the same color as him/her. The actual value of $k$ was randomly selected in each run of the experiment (see  Materials and Methods below and  \nameref{setup} and \nameref{beepingcases} for further details about the tracking system and the experimental procedure). Knowing the real-time positions of the subjects, albeit with a typical uncertainty of 30\,cm, the computer   determines the $k$ nearest neighbors of each individual, and instructs her/his left tag to beep if the majority of these neighbors are {not}  the same color as hers/his. In the final ``silent'' state, when nobody beeps anymore, a given value of $k$  imposes connected groups of at least $(k+1)/2+1$ nearest neighbors of the same color (see \nameref{fig:1}c for $k=5$ and  \nameref{final_config_1}), so that one could expect  a priori that the mean group size should grow with $k$.

Our experimental setup allocates to each individual an \textit{artificial sensory device} allowing the probing of their environment. Although we have a perfect control of the meaning of the signal received by each individual through the selection of $k$, the subjects have only a very vague and incomplete understanding of its nature (again, we only mentioned the notion of ``environment'' to them, and never explained its precise definition). In addition, for a given $k$, this sensory device has some inherent limitations as it only probes a finite range of the environment (the $k$ nearest neighbors) and only returns the nature of the dominant color within this range (by beeping or not beeping).  In a sense, this is similar to the human eye, a limited sensory device which can only identify objects up to a certain distance  and which can only capture a finite range of the light spectrum. However, in our experiment, the signal itself is simple and binary (beep/no beep), thus avoiding any cognitive saturation phenomenon. As will be shown hereafter, selecting the value of $k$ allows to control the structure of the final state, and in particular the quality of the final segregation, {i.e.}, the number of formed groups of each color and the possibility to attain a fully separated state with only two well-defined groups of 11 red and 11 blue subjects (see  \nameref{final_config_1}).

\paragraph*{Fig~4}
{\small\textbf{Representative selection of final states for all values of ${k}$}. The quality of the separation increases (i.e. less groups are formed in the final state) with $k$ up to $k=7$, value above which it saturates.}
\label{final_config_1}

\subsection*{Separation dynamics into unicolor groups}
Quite remarkably, all experimental runs with all values of $k$ ultimately resulted in a silent final state (see \nameref{fig:1}a bottom panel and  \nameref{final_config_1}, and  \nameref{S1_video} and  \nameref{S2_video}), where the subjects were also found to  remain still. 
This is illustrated in \nameref{fig:1}b, where the fraction of beeping individuals averaged over all runs is plotted as a function of time. For all $k$, after a short transient time of $\sim 10$\,s, we find that this fraction of beeping subjects displays an approximate exponential decay (see  \nameref{NumberGroupslog}). During the segregation phase, we observe that the subjects spontaneously stop walking when their tag is not beeping, and resume walking if their tag beeps again, although no  instructions were ever given to them regarding how to behave with respect to the state of their tag (see the verbatim instructions in the experimental protocol below).

\paragraph*{Fig~5.}
{\small\textbf{Instantaneous fraction of beeps for each value of ${k}$}:
	Purple lines show the experimental data and red lines model simulations.
	The y-axis is in log scale, evidencing the near-exponential decay.} \label{NumberGroupslog}\vskip 0.4cm

In order to compare the dynamics and the final state for different values of $k$, we  define the notion of 3-group, independently of $k$: at any time, and in particular in the final silent state,  we connect each individual to the neighbors of the same color among her/his 3 closest neighbors. The 3-groups are then defined as the connected components in the resulting graph. The notion of 3-group is quite natural when seen as the dual structure associated to a Delaunay triangulation or in relation with the 3-vertex structure of a Voronoi planar lattice~\cite{geom}. In addition, it certainly makes sense to decorrelate the analysis  metric (here, 3-groups) and the information generation metric ($k$-neighbors), because all experimental runs appear to be of the same nature from the subjects' perspective (whatever $k$). 
In practice, we find that this precise and $k$-independent definition leads to intuitive group configurations (see the bottom panel of \nameref{fig:1}a  and  \nameref{final_config_1}). \nameref{fig:1}b shows that the mean number of (3-)groups decays after a  transient time already observed for the fraction of beeps, and ultimately saturates in the final state.

\subsection*{Model for the random walk phase}
In order to understand these results along with additional characterizations of the final state for the different values of $k$, we introduce a data-driven model for the motion of the subjects which will be described in more details in the following  Materials and Methods section (see in particular Eq~(\ref{eqmain}--\ref{interfh})). The model describes the repulsive interaction between agents, and between each agent and the boundary of the arena. These interactions  take into account the human anisotropic perception of the physical environment, which leads to a stronger avoidance when an obstacle is in front of an agent than behind it. In addition, the model includes a friction term  that tends to give a typical \textit{target speed} $v_0\sim 1$\,m/s to the agent, as well as a \textit{stochastic noise} of intensity $\sigma_0$ describing the spontaneous fluctuations in the motion of the agent.

\paragraph*{Fig~6.}
{\label{fig:2}\small\textbf{Quantitative analysis of the model for 22 subjects in the random walk phase.} The experimental and  the model results are respectively plotted in black and red. (\textbf{a})~Probability distribution function (PDF) of the instantaneous speed of the subjects (agents in the model).   (\textbf{b})~PDF of the distance between a subject/agent and its closest neighbor at a given time. (\textbf{c})~PDF of the distance of the subjects/agents from the center of the arena. (\textbf{d})~PDF of the relative angle between the velocity of a subject/agent and the normal to the boundary of the circular arena. The raw data (blue dashed line) show that the subjects in the experiment turn more often anticlockwise, keeping the boundary of the arena to their right. The black curve is obtained by symmetrizing the data, i.e., by adding all mirror trajectories to the data set. The agent model does not have any intrinsic left/right asymmetry. (\textbf{e})~Time correlation function of the velocity $C_v(t)=\langle\vec{v}(t+t')\vec{v}(t')\rangle$, where $\langle\,{\rm \bullet}\,\rangle$ stands for the average over all experimental runs and over the reference time $t'$. Oscillations translate the velocity anti-correlation developing due to the bounded arena. (\textbf{f})~Mean square displacement of a subject/agent $D_x(t)=\langle[\vec{x}(t+t')-\vec{x}(t')]^2\rangle$. For large time, when the trajectories decorrelate, it ultimately saturates to $2\langle\vec{x}^2(t')\rangle$, i.e., twice the mean square distance between a subject/agent and the center of the arena (at $\vec{C}=(0,0))$, which can also be evaluated from (\textbf{c}).}\vskip 0.4cm

In \nameref{fig:2}, we compare the results of the model with 22 agents to experimental data  obtained during the random walk phase before the beeps are switched on, of typical duration of 45\,s in each experimental run. The model predictions for  the probability distribution functions (PDF) of (a) the speed, (b)  the distance to the closest neighbor, (c)  the distance to the center of the arena, and (d) the relative angle of the velocity to the normal to the boundary of the arena are in good agreement with the experimental results. In addition, the velocity correlation function and the mean square displacement as a function of time -- two intrinsically dynamical quantities --  are also well reproduced by the model.

\subsection*{Model for the segregation phase}

When the beeps are switched on, we propose a very simple \textit{strategy} for the agents controlled by our model: if an agent is \textit{not} beeping (i.e., if the majority of its $k$ nearest neighbors is of the same color as that of the agent), we set the target speed  and the stochastic noise to zero, so that the agent will ultimately stop. However, whenever the agent beeps,  we set the target speed to $v_{\rm beep}$ and also switch on the stochastic noise of intensity $\sigma_0$ (the latter at the same value as in the random walk phase; see the detailed description of the model in Materials and Methods, and in particular, Eq~(\ref{eqmain}--\ref{eqfin0})). The target speed $v_{\rm beep}\sim 0.4$\,m/s is  found to be smaller than in the random walk phase (see \nameref{Table2} and \nameref{Table3}). Indeed, because the subjects are paying attention to the sound emitted by  their left tag, they tend to move slower than in the random  walk phase. Note that some human subjects appear to exploit the additional strategy of visiting/``sniffing'' already formed groups of immobile subjects  to check if they are beeping while standing near such a group. This strategy, which is only relevant in the late stage of the segregation dynamics when enough stable groups are already formed, is not implemented in our simple model.

The model qualitatively reproduces  the separation dynamics (see   \nameref{S3_video}) and quantitatively recovers  the exponential-like decay of the fraction of beeps and of the number of groups with time (\nameref{fig:1}b and  \nameref{NumberGroupslog}), as well as other quantities like the distribution of individual stopping times (the time until a subject has stopped beeping for good) and the distribution of the total time during which each individual has been beeping (see  \nameref{time-PDFs}).

\paragraph*{Fig~7.}
{\small\textbf{Probability density functions (PDF) of the normalized segregation time}.
	(a) $\bar{t}_{\rm b} = t_{\rm b}/\left< t_{\rm b}\right>$, where $t_{\rm b}$ is the total
	time a given individual spends beeping.
	(b) $\bar{t}_{\rm f} = t_{\rm f}/\left< t_{\rm f}\right>$, where $t_{\rm f}$ is the final
	time a given individual has beeped.
	(c) $\bar{t}_{\rm end} = (t_{\rm end} - \left< t_{\rm end}\right>)/\sigma_{t_{\rm end}}$,
	where $t_{\rm end}$ is the maximum value of the  $t_{\rm f}$ for a given experimental run, and hence coincides with the total duration of this run.
	Black lines: experimental data; red lines: model simulations. The blue line in (c) is the
	universal Gumbel distribution characterizing the distribution of the maximum of independent (or weakly dependent) random variables ($t_{\rm end}$ is the maximum of the individual ${t}_{\rm f}$).
	In (a) and (b), mean values $\left< \bullet \right>$ take into account all pedestrians from
	all sessions and all values for $k$; in (c), the mean value is calculated over all sessions
	and all values for $k$.}
\label{time-PDFs}

\subsection*{Characterization of the final state}
We now focus on the properties of the final silent state, where the subjects are not beeping anymore and ultimately stop moving. For the seven values of $k$ investigated experimentally, \nameref{fig:3} shows the final distribution of the size of the 3-groups  and  \nameref{final_config_1} shows a representative selection of the spatial organization of subjects in the final state. For $k\leq 5$, and in particular for $k=1$, we observe that the final state is often far from being perfectly segregated (i.e., exactly two groups of 11 subjects), with a non negligible probability to observe groups of less than 11 individuals (75\,\% cumulated chance to observe groups of size 2 to 9 for $k=1$). However, for $k\geq 7$, we find that the distribution of group sizes does not evolve much. This can be also seen in the probability of occurrence of groups of size 11, which increases from its minimal value 0.25 for $k=1$, before saturating around 0.8 for $k\geq 7$. 

\paragraph*{Fig~8.}
{\small\label{fig:3}\textbf{Characterization of the unicolor groups in the final silent state.} The experimental and  the model results are respectively plotted in black and red (the model plots are slightly shifted to the right for better readability). Probability distribution function (PDF) of the size of the 3-groups in the final silent state are shown for $k=1,3,5,7,9,11,13$. The vertical scale is adapted to focus on the probabilities to find a group of size strictly less than 11. Note that due to the experimental noise in the positions of the subjects, a few groups of size 1 and 2 were experimentally found in some instances. The last panel on the bottom right summarizes the results using a larger vertical scale to reveal the increase and the ultimate saturation of the fraction of groups of size $22/2=11$.}\vskip 0.3cm

\paragraph*{Fig~9.}
{\label{fig:4}\small\textbf{Characterization of the efficiency of the separation as a function of $k$.} The experimental and  the model results are respectively plotted in black and red. The blue symbols and lines correspond to the results of a series of experiments where the subjects were additionally instructed to form  two clearly identifiable groups in the final state. (\textbf{a})~Average size of the 3-groups in the final silent state, as a function of $k$.  (\textbf{b})~Average number of groups. (\textbf{c})~Fraction of groups of maximum size 11. (\textbf{d})~Probability to observe a perfect separation in two groups of size 11. All panels illustrate the improvement of the segregation in the main experiment and the model up to $k=7$, above which it saturates. For the experiment with the additional instruction of forming two clearly identifiable groups, the efficiency of the segregation is much improved even for small $k$, and only slightly increases with $k$ (blue circles and line).}\vskip 0.4cm

\nameref{fig:4} shows the resulting mean group size as a function of $k$, which first sharply increases from the value 6.5 for $k=1 $, before saturating around 9.5-10 for $k\geq 7$.  Accordingly, the mean number of groups decreases from 3.4 for $k=1 $ and then saturates around  2.2 for $k\geq 7$. 
\nameref{fig:4} also recapitulates the probability of occurrence of groups of size 11 and the probability to observe a perfect segregation. Both quantities again illustrate how the separation quality only improves  up to $k=k_{\rm sat}=7$, the probability to observe a perfect separation saturating to around 0.8 for $k\geq 7$. The results of the model presented in \nameref{fig:3} and \nameref{fig:4} are in good agreement with experiments. 

The model also allows us to explore how our results would depend on the number  $N$ of pedestrians ($N=22$ in the experiment) and on the pedestrian density ($D=N/(\pi R^2)=0.55$ pedestrian/m$^2$ in the experiment). 
In particular, the value $k_{\rm sat}$ above  which the number of groups and the fraction of groups of size $N/2$ saturate is   found to be essentially independent of the density, and $k_{\rm sat}$  grows linearly with the number of pedestrians $k_{\rm sat}\approx N/3$ (see  \nameref{SaturationSI}). 
This linear relation can be understood by the fact that for a given value of $k$, groups of size $(k+1)/2+1\sim k/2$ are stabilized. If there are only 3 unicolor groups (near perfect separation), the smaller group is necessarily of size smaller than $N/4$ and hence $k_{\rm sat}/2<N/4$, which leads to $k_{\rm sat}<N/2$.

\paragraph*{Fig~10.}
{\small\textbf{Saturation analysis for different group sizes and densities}: Values of $k$ above which the average number of groups -- and other related quantities (see \nameref{fig:4}) -- saturates, for different (a) group sizes $N$ and (b) pedestrian densities $D'$ with $N = 22$ agents (like in the experiment; $D$ is the actual experimental density of 0.55 pedestrian/m$^2$), and the corresponding saturation values of the average number of groups (c and d). $k_{\rm sat}$ is found to grow like $\approx N/3$ (red dashed line in a) and to be essentially independent of the density (red dashed line in b).}
\label{SaturationSI}

\subsection*{Segregation in two clearly identifiable groups}
In other experiments conducted after the main series, in order to avoid influencing the participants in the main experiments, the subjects were additionally instructed to form a final state with \textit{two clearly identifiable} groups (see  \nameref{final_config_2} and  \nameref{S4_video}). This additional instruction resulted in shorter segregation times and sharply improved the quality of the separation for all values of $k$ (see \nameref{fig:4}; blue lines and symbols). Interestingly, most participants probably had the intuition that ultimately regrouping in two distinct groups would facilitate the color separation process. Yet, it was only once they were specifically instructed to do so, and hence when they knew that the other subjects shared this same strategy, that the group was indeed able to exploit this strategy.  In a sense, this phenomenon is reminiscent of certain forms of ``prisoners and hats" logic puzzles \cite{hat} where a group is able to collectively find the solution of a problem only after each individual is aware that the others will share the same strategy as them.

\paragraph*{Fig~11.}
{\small\textbf{Representative selection of final states for all values of $k$}, in the experiments where subjects were explicitly asked to separate in two distinct groups.
	The separation was then perfect in most cases.}
\label{final_config_2}

\subsection*{Analogy with a physical phase separation}
The present experiment also has  a surprising analogy in the physical context, being formally similar to the phase separation of two materials at very low (formally, zero) temperature, forcing the system to only decrease its energy during its evolution. For simplicity, let us define a toy version of our experiment on a one-dimensional lattice where all sites are initially randomly occupied by a blue or red agent (two immiscible phases like water and oil in the physical context). We associate an energy cost to each neighboring pair of agents of different colors (surface tension). The dynamics proceeds by the exchange of the positions of neighboring pairs of distinct colors which would decrease or would not change the energy (zero temperature), {i.e.}, that would decrease or would not change the number of heterogeneous contacts (surface energy). It is easy to see that an agent of a given color will exchange position with one of its neighbors only if both neighbors have the other color. In a sense, in the context of our  experiment, this problem is somewhat associated to $k=2$, where an agent ``beeps'' and moves only if the strict majority of its 2 neighbors has a different color. The dynamics  hence progressively eliminates isolated individuals but will ultimately freeze in an incompletely separated final state, because all subsequent exchange moves would increase the energy (i.e., the number of heterogeneous contacts), which is forbidden at zero temperature. This problem was analytically solved in \cite{majsire}, where the fraction of ``beeping'' agents was  asymptotically shown to decay exponentially, as observed in the present experiments and associated model (see  \nameref{NumberGroupslog}),  and where the final state was characterized exactly: number of groups or interfaces, average size of the groups, distribution of group sizes... Note that the relevance in the social context of the analog of phase separation was already considered in \cite{schelling1971,schelling_micromotives_1978}, where models directly inspired from physics are exploited.

\section*{Discussion}

Understanding how interactions between individuals determine the collective behavior of animal  and human groups is a fundamental issue in the natural and social sciences \cite{castellano_statistical_2009,schelling_micromotives_1978,camazine_self_2001,goldstone_collective_2009,sumpter_collective_2010,helbing_social_2012}.
Not only the way in which individuals exchange information, but also the amount of information collected by an individual about  the behavior of the other group members play a key role in the collective forms of organization that emerge at the group level and  in the capacity of a group to provide efficient responses to perturbations \cite{moussaid_collective_2009,camperi_spatially_2012,wang_how_2018}. 
Today, because of the massive use of digital interfaces mediating the interactions between individuals in human societies, it is crucial to understand the impact of information filtering on the collective behavior and on the level of polarization of groups \cite{pariser_filter_2011,flaxman_filter_2016}. In most cases, the information exchanged between individuals remains limited in comparison to the size of the group and each individual only interacts with a limited number of  group members.

We have presented a series of experiments where a group of subjects equipped with an artificial sensory device had to perform a collective task, attempting to achieve a more or less complete (phase) separation.  This fully controllable artificial sensory device filters and adapts in real time the information that is delivered to the participants, so that their decision to stop or resume walking only depended on whether or not the majority of their neighbors was of the same color as them. The participants being unaware of the actual nature of this control allowed us to investigate the impact on the separation process and final state of the amount of information available to the subjects about their environment (i.e., the perception range $k$ of the sensory device). 

We have shown that the amount of information collected at the individual scale, controlled by the range $k$, deeply affects the collective separation dynamics and the resulting final  state observed at the group level. Our results show that there is a minimal level of information required to get an optimal level of segregation ($k_{\rm sat}= 7$), beyond which no increase in performance is observed. In particular, the dynamics (decay of the fraction of beeping individuals and of the number of groups, individual beeping periods...) and the structure of the final state (distribution of group sizes, probability of a perfect separation...) were fully characterized and are well reproduced by a data-driven model which also quantitatively describes the dynamics of freely walking humans in a confined space. The model demonstrates that the simple and resilient strategy -- in practice adopted by the actual human subjects -- to walk and stop, depending on whether an individual ``beeps'' or not, always resulted in a final silent state where each individual had a majority of its $k$ nearest neighbors of the same color. The model also permits to explore other conditions (number $N$ of subjects, subject density, values of $k$...) not investigated experimentally, revealing in particular the linear dependence with $N$ of the value of $k$ for which no improvement in the segregation is observed, $k_{\rm sat}\sim N/3$.
We also established a strong analogy between our experiments and the actual physical process of  phase separation at zero temperature, where the system can be ultimately frozen in a metastable (local minimum of energy) and incompletely separated final state, due to the rapid cooling (quench). 
In addition, similar experiments where the subjects were also instructed to produce a final state with two clearly identifiable groups resulted in much better separated final states, illustrating that a group can achieve a much better performance when all individuals are aware that they share the same strategy, a phenomenon reminiscent of a class of ``prisoners and hats''  problems.

With the sensory interface used in the experiments, the individuals were able to acquire and process information about their environment without cognitive limitation. There was no information overload since the performance of the sensory interface remains the same regardless of the amount of information (small or large $k$) that is gathered by the individuals about their environment, the information being channeled into a single-bit beep. These segregation experiments have shown that there exists a maximum amount of information beyond which the acquisition of additional information at the individual level does not improve the performance of the group and that this limitation is not the consequence of an overload of the information processing system. These experiments also illustrate how the choice of the information delivered to a human group (here, by selecting the value of $k$) allows to \textit{control} the dynamics and the resulting final state of the group, and in particular, the typical size of unicolor groups, even when the individuals are not  aware of the precise nature of the information externally provided.  This result echoes the issue of controlling the excessive polarization of human groups in many social contexts, in particular due to information filters~\cite{bozdag2015}. 

Finally, in the  general context of collective animal behavior studies, our results bring new insights regarding the amount of information that each individual in a group needs to collect on their neighbors in order to get an efficient coordination  at the collective level. For instance, in the case of starling flocks, it has been suggested that each bird interacts on average with a bounded number of neighbors (typically 7) rather than with all neighbors within a fixed metric distance~\cite{Ballerini2008}. One may wonder if the existence of such limited amount of information gathered by individuals is only fixed by the cognitive abilities of the considered species to process that information. Indeed,  if the information collected by an individual exceeds their cognitive abilities, this would lead to some kind of information overload~\cite{klingberg_overflowing_2009,saegert_crowding_1973}. Our results suggest that, in the case of the simple spatial segregation of a human group, there is a maximum amount of information beyond which the acquisition of additional information by individuals does not improve the group performance. Moreover, our experiments demonstrate that this limitation is not the mere consequence of cognitive overload.

\section*{Materials and Methods}

\subsection*{Experiments}
\subsubsection*{Participants and ethics statement}
Two series of experiments were conducted in September 2015 and June 2016 at  the Universit\'e Paul Sabatier in Toulouse, France.
A total of 115 participants in 2015 and 209 participants
in 2016 took part in the study. They were naive to the purpose of our experiments and
gave written and informed consent to the experimental procedure and to the use of pictures and movies of the experiments featuring them. Each participant was
paid 10\,{€}  per hour, whatever the number of sessions he or she participated in.
The participants had to walk in closed circular arenas according to predefined rules
that were given at the beginning of each series of experiments.
The aims and procedures of the experiments have been approved by the Institutional
Review Board of Inserm (IRB00003888, decision n$^\circ$15-243).

\subsubsection*{Experimental setup}
The experiments were performed in a large hall where circular arenas of radius 1.78\,m ($\approx 10$\,${\rm m}^2$; yellow tape), 2.52\,m ($\approx 20$\,${\rm m}^2$; red tape) and 3.56\,m ($\approx 40$\,${\rm m}^2$; blue tape) were marked on the ground (see  \nameref{setup}). 
Only the largest arena (blue tape) was used in the sessions with 22 individuals. The two other arena sizes were used to build a model of pedestrian motion in confined arenas.
Individuals' trajectories were tracked using a real-time location system developed
by Ubisense and based on Ultra-Wide Band (UWB) signals triangulation \cite{souk_2016}. When compared to
camera-based tracking of individuals, the fundamental feature of UWB-based localization
is that the tracking of each individual in a group is 100\,\% accurate: each individual is
assigned a uniquely defined pair of tags that are unambiguously associated with the
localization of the individual.

The tracking system consisted  of pairs of tags clipped above the participant's
shoulders, that emitted UWB wave trains, and sensors precisely placed in
order to cover the experimental arena, that received and processed signals from tags \cite{souk_2016}.
In our experiments, the tracking system included 6~to~8 sensors uniformly distributed
around the arena, fixed 2-3\,m from the border of the arena and 4\,m from the ground.
Ubisense sensors, as depicted in  \nameref{setup}, are UWB signal receivers linked
together and to the server by high-speed low-latency Ethernet connections.
These devices are rectangular boxes of size $21.5 \times 15 \times 9$\,cm and weight
1\,kg. All these sensors were wire-connected through a router to a server that actually
performed the localization of tags and was able to send back some information to the
tags.

Ubisense tags are miniaturized circuits powered with batteries that operate both in the
6~to~8\,GHz frequency band (UWB signals used for localization) and in the 2.4\,GHz band
(signals used for data exchange and synchronization). Each participant was wearing two
tags, one on each shoulder ($8.3 \times 4.2 \times 1.1$\,cm and 32\,g for the left
tag; $3.8 \times 3.9 \times 1.65$\,cm and 25\,g for the right tag), attached by clips.
These tags can emit an acoustic signal (a beep) triggered by the central server that has
a global view of all participants’ locations and colors.
Sensors were tightly synchronized together to measure the positions of tags. Each sensor
captured the electromagnetic wave signals (pulses) emitted by the tags and measured in
real time the angle from each tag to the sensor with an angular accuracy of 1.5$^{\circ}$.
Additionally, pairs of sensors computed the time difference of arrival for a given tag
pulse train. The server received, through a wired Ethernet connection, all available information
(angles and time differences of arrival for each tag emission) and was able to derive the
tags positions. 

In our experiments, the information collected by the sensors was used to evaluate in real
time the environmental conditions of each subject by processing the relative position
and color of all the subjects, and to react according to defined experimental conditions (the value of $k$).
These conditions (see  \nameref{beepingcases} and \nameref{Table1}) determine the state of a tag, that is, whether it
should emit an acoustic signal or not. The information about the acoustic signal
(beeping or silent) was sent back to each individual with a frequency of 1\,Hz so that
individuals could react to this information in real time with a short delay of less than
one second. 
The acoustic signal emitted by a tag had a short duration and low intensity and did not interfere with the sound of the other participants' tags. 
The system was set up to perform the real-time double tracking (two tags per individual) of 22 individuals, whose
instantaneous positions were determined with an error of less than 30\,cm and recorded
with a typical  frequency of 2\,Hz for each tag. Moreover, due
to the time-sharing, tags did not emit signals at exactly 2\,Hz, although our system
was accurate and the communication flux was highly homogeneous.  \nameref{tag-freq} shows
that the disparity within a session is also quite small.

In addition to the tracking system, the
experiments were recorded with two cameras, a 360$^{\circ}$ camera fixed to the ceiling of
the experimental room, and a camera located at 4\,m  from the floor, providing a
lateral view. 

Note that the typical segregation time was found to be approximately twice longer in the experiments from June 2016 compared to the experiments from September 2015, probably due to a slightly smaller  tag recording frequency in the 2016 experiments.
However, this issue only alters the segregation speed and hence its dynamics and not the properties of the final  state, as also supported by our model when changing time scales.
We compared the results in the final state for both experiments separately, and indeed found  no significant difference between them. We could thus combine all data for the final state analysis.
Yet, in \nameref{fig:1}b and  \nameref{NumberGroupslog}, we show the dynamics for the September 2015 experiments only, which excludes the case $k=13$. The dynamics for June 2016 are equivalent but with an exponential decay about twice slower. 
They are also much  noisier (and hence not worth showing) for $k = 1$ to 11, because only about 5 experiments were run in each case (see  \nameref{Table1}).

\subsubsection*{Experimental protocol}
In a first series of experiments, called ``classic segregation'', participants were not
explicitly informed that they had to segregate in two groups. The participants received
the following instructions at the beginning of each one hour session: 
\begin{quote}
	\begin{enumerate}
		{\it 
			\item Each  of you has been assigned one of two colors, red or blue. None of you
			knows her/his own color or the color of the others. Your color may change from one
			experimental run to another. You are asked not to communicate in any way during the experiment.
			\item At the beginning of an experimental run, you will walk randomly at your normal pace, remaining in the
			arena, and avoiding colliding with other participants.
			\item After this random walk period of typically 45\,s, your left tag will be switched on and will start to emit a beep whenever your environment is not the same color as you.
			\item The experiment will stop when all tags will be silent ({\it nobody beeps}).
		}
	\end{enumerate}
\end{quote}

Once all the sessions of classic segregation had been carried out, a second series of
experiments, called ``segregation in 2 clusters", was performed in which the participants
were explicitly instructed to separate in two clearly identifiable clusters.
The fourth rule became:
\begin{quote}
	\begin{enumerate}
		{\it 
			\item[4'.] The experiment will stop when all tags will be silent ({\it nobody
				beeps}), and two clusters will be clearly identifiable.
		}
	\end{enumerate}
\end{quote}

The ``segregation in 2 clusters'' experiments were performed after all ``classic segregation'' experiments had been carried out, because the instruction of
``forming two clusters'' was a key indication that could have artificially biased the performance of the groups in the latter case.

At the beginning of each session, participants were randomly assigned a color
(11~red and 11~blue) by the server and were instructed to walk randomly within the arena at
their comfort speed,  not being allowed to have any oral or facial
communication with each other. Moreover, participants were not informed that colored
groups were of equal size. 

\nameref{beepingcases} illustrates the rule governing the emission of a beep by a tag: 
the tag carried by a participant emits an acoustic signal if the color of the majority of his/her $k$ nearest neighbors is different from his/her own color. 
We used odd values for $k$ ($k$ = 1, 3, 5, 7, 9, 11, and~13), in order to have a proper
definition of the majority. 
Overall, 303 sessions were carried out (see  \nameref{Table1}). For the different values of $k$, some typical final configurations are presented in  \nameref{final_config_1} (``classical segregation'') and \nameref{final_config_2} (``segregation in 2 clusters'').

\subsection*{Data Collection and preprocessing}

\subsubsection*{Data structure}

The data  collected by the system for an individual in a given session consist of two
series of triplets, $\{t^{\rm L}_i,x^{\rm L}_i,y^{\rm L}_i\}_{i=1}^{N_{\rm L}}$ and
$\{t^{\rm R}_j,x^{\rm R}_j,y^{\rm R}_j\}_{j=1}^{N_{\rm R}}$, which correspond to the left
and right tags respectively, where $(x_i,y_i)$ is the spatial position of the tag at a given
time $t_i$. Typically, both time series are not of the same length
($N^{\rm L} \ne N^{\rm R}$), they are not synchronized
($|t^{\rm L}_i-t^{\rm R}_j| > \varepsilon$ for $i=1,\dots,N^{\rm L}$ and
$j=1,\dots,N^{\rm R}$, with $\varepsilon \approx 0.2$~s) and, although they are
quite regular
($t^{\rm L}_{i+1}-t^{\rm L}_i \approx t^{\rm R}_{j+1}-t^{\rm R}_j \approx 0.5$~s), there
are some gaps or missing data at different positions in each series.

Besides these usual pathologies in data processing, we also detected an artificial
temporal non-uniformity in the arrival of data streams, as if the data arrived in waves.
Our guess is that the TCP/IP protocol accumulates the data collected in the router from
the different sensors, and awaits for data packets to be sufficiently large before
transmitting them to the computer. As time is defined by the arrival to the
database, an artificial delay is generated in the time series, giving rise to an
undesirable ``burst'' effect in the dynamics of a normal walking pace.

In order to get rid of these bursts, we designed a procedure (described in the next paragraph) specifically aimed at smoothing the trajectories. We call the new trajectories ``reconstructed'' (see example of original and reconstructed trajectories in  \nameref{traj-reconstr}).

%%%%%%%%%%%%%%%%%%%%%%%%%%%%%%%%%%%%%%%%%%%%%%%%%%%%%%%%%%%%%%%%%%%%%%%%%%%%%%%%%%%%%%%%%
\subsubsection*{Data preprocessing: synchronization and time rectification} \label{time_rectif}

Ideally, the information provided by the two tags of an individual should determine both
his/her position and orientation, if it is assumed that shoulders are always perpendicular to the direction of motion. This is however
not the case: people tend to rotate their body when crossing or approaching each other,
or even with no apparent reason (see the random walking phase in the movies). Therefore, we  integrated the information from  both tags in a single data and described individuals by a single point located in the geometric center
of the segment defined by the two tags. The error in the location of this point is one
half of the error in the location of the shoulders (i.e., $\approx 15$\,cm). We then considered that the
orientation of the individual is determined by the direction of the velocity vector.

Our analysis required the knowledge of the position and color of each of the $N=22$ individuals simultaneously,
at each time instant.
The data were synchronized in a common time scale, and in order to allow spatial derivation to
estimate the velocity and acceleration, we used a small time step,
$\Delta t = 0.1$\,s, again by linear interpolation, keeping in mind, during the posterior
statistical analysis, that {\it real} data had a time step of  0.5\,s.

Finally, we corrected the ``burst'' effect by assuming that positions were well
calculated and that time steps had to simply be redistributed in the time series
according to a homogeneous density along the series (see  \nameref{datawaves}).

The detailed steps of the data reconstruction procedure are:
\begin{enumerate}
	\item Synchronization of the times series for both tags:
	each time series is first supplemented by linear interpolation in order to have the same time instants for both tags. 
	Both time series were merged in increasing order in a
	single time series $\{\hat{t}_i\}_{i=1}^{N_{\rm H}}$ of length
	$N_{\rm H} = N_{\rm L}+N_{\rm R}$. The position of each respective tag at the new
	instants of time, $\{\hat{x}^{\rm L}_i,\hat{y}^{\rm L}_i\}_{i=1}^{N_{\rm H}}$ and
	$\{\hat{x}^{\rm R}_i,\hat{y}^{\rm R}_i\}_{i=1}^{N_{\rm H}}$, were calculated by linear
	interpolation of $\{x^{\rm L}_i,y^{\rm L}_i\}_{i=1}^{N_{\rm L}}$ and
	$\{x^{\rm R}_j,y^{\rm R}_j\}_{j=1}^{N_{\rm R}}$ respectively, at times $\hat{t}_i$.
	Then, $\Delta \hat{t}_i = \hat{t}_{i+1}-\hat{t}_i$ was approximately equal to 0.2
	and 0.3 s, alternatively.
	
	\item The position of the pedestrian at a given time was then calculated by linear
	interpolation of the position of the two tags:
	$x^{\rm H}_i = (\hat{x}^{\rm L}_i + \hat{x}^{\rm R}_i)/2$,
	$y^{\rm H}_i = (\hat{y}^{\rm L}_i + \hat{y}^{\rm R}_i)/2$, for $i=1,\dots,N_{\rm H}$.
	
	\item Time instants were redistributed according to a locally averaged velocity.
	The new time series $\{s_i\}_{i=1}^{N_t}$ was built recursively as follows:
	\begin{align}
	s_1 & = \hat{t}_1, \\
	s_{i+1} & = s_i + \lambda \frac{d_i }{ \bar{v}_i},
	\quad i=1,\dots,N_t-1,
	\label{eq2}
	\end{align}
	where $d_i = \| \vec{u}_{i+1} - \vec{u}_i \|$, $\vec{u}_i = (x^{\rm H}_i,y^{\rm H}_i)$,
	and $\bar{v}_i$ is an averaged velocity:
	\begin{align}
	\bar{v}_i = {
		\displaystyle{\sum_{k=-\infty}^{+\infty} }
		\| \vec{u}_{i+1+k} - \vec{u}_{i+k} \| \; e^{-(\hat{t}_{i+k}-\hat{t}_i)^2/t_c^2}
		\over 
		\displaystyle{\sum_{k=-\infty}^{+\infty} }
		( \hat{t}_{i+1+k} - \hat{t}_{i+k} ) \; e^{-(\hat{t}_{i+k}-\hat{t}_i)^2/t_c^2}},
	\label{decadix}
	\end{align}
	with $t_c$ the ``radius'' of the time interval centered on $\hat{t}_i$ in which the
	exponential has a significant value, and $\lambda$ a normalization constant ensuring
	that the total duration is preserved: $s_N - s_1 = \hat{t}_N - \hat{t}_1$.
	Then:
	\begin{align}
	s_N - s_1 & = s_N - s_{N-1} + s_{N-1} - s_{N-2} + s_{N-2} - \dots \nonumber\\
	& - s_2 + s_2 - s_1\\
	& = \lambda \left( {d_{N-1} \over \bar{v}_{N-1}} + {d_{N-2} \over \bar{v}_{N-2}}
	+ \dots + {d_1 \over \bar{v}_1} \right)
	\\ &= \lambda \sum_{i=1}^{N-1} {d_i \over \bar{v}_i}
	= \hat{t}_N - \hat{t}_1, 
	%\\
	%& \hspace{2.5cm} \mbox{so} \quad 
	%\lambda = { \hat{t}_N - \hat{t}_1 \over \displaystyle{\sum_{i=1}^{N-1} {d_i \over \bar{v}_i}}}.
	\end{align}
	i.e., $\lambda = (\hat{t}_N - \hat{t}_1) \times
	\displaystyle{ \left(\sum_{i=1}^{N-1} {d_i \over \bar{v}_i} \right)^{-1}}$.
	%Note also that
	%$\, s_{i+1} = s_1 + \lambda \displaystyle{\sum_{j=1}^i {d_j \over \bar{v}_j}}$,
	%\, so
	%$\, s_N = s_1 + \lambda \displaystyle{\sum_{i=1}^{N-1} {d_i \over \bar{v}_i}}$.
	
	If $t_c=0$, then only the term with $k=0$ remains in Eq~(\ref{decadix}), and
	$\bar{v}_i$ becomes:
	\begin{align}
	\bar{v}_i = {\| \vec{u}_{i+1} - \vec{u}_i \| \over\hat{t}_{i+1} - \hat{t}_i }
	= {d_i \over \hat{t}_{i+1} - \hat{t}_i }.
	\end{align}
	Thus, for $\lambda=1$, Eq~(\ref{eq2}) gives: $s_{i+1} - s_i =  \hat{t}_{i+1} - \hat{t}_i$, from which we get $s_i = \hat{t}_i$, i.e., that time instants are not redistributed.
\end{enumerate}

The critical parameter $t_c$ is the radius of the time window over which the velocity
is averaged. The choice of the value for  this parameter is based on making a compromise between reducing
the impact of the bursts on the normal pace, and preserving the speed variation inherent
to a normal walk.

\nameref{datawaves} shows the successive positions (circles) of a random individual,
during a portion of her/his trajectory when walking randomly (upper panels) and during a
segregation phase (lower panels). The circles are equally spaced in time so that the bursts
appear clearly as dark rings (no spatial variation), especially for small values of $t_c$ and in the segregation
phase, where the speed is in general smaller than in the random walk phase.

For larger values of $t_c$, the circles are homogeneously redistributed along the
trajectory: for $t_c=0.5$\,s in the upper panels and $t_c = 1$\,s in the lower ones, the
bursts have disappeared, but with the inconvenient loss of diversity in the walking
speed, whose variation has been practically removed.

An exhaustive trial and error analysis showed that the best compromise was reached
when $t_c=0.4$\,s. 

We have also performed an analysis of the spatial error introduced by the redistribution
of time instants, by evaluating the spatial shift that individual positions undergo
when $t_c > 0$ with respect to their location when $t_c=0$~s. Note that the position
at time $t$ when $t_c=0$~s is not necessarily the true one, which can only be evaluated
by comparing it with the videos of the direct experiments, meaning that this analysis is only a
measure of the spatial shift produced by the use of different values of $t_c$.

\nameref{decalage} shows that the spatial shift is almost always smaller than 40~cm,
which is the typical width of an individual (distance between shoulders).

\subsection*{Model}
Here, we briefly describe the model used to simulate the initial random walk phase and the subsequent segregation phase leading to the final silent state. This model is a simpler version of a data-driven model derived by the same authors, which will presented elsewhere, exploiting the processing (\nameref{tag-freq}--\nameref{decalage}) and analysis of many 3 minutes one-subject experimental trajectories (to infer the spontaneous motion of a subject and their interaction with the boundary of the arena) and two-subjects trajectories (to infer the repulsive interaction between subjects). This methodology was introduced in \cite{fishint} and  applied successfully to the determination of the social interactions in a fish species and its detailed and very similar  implementation in the context of human groups will be presented elsewhere.

\paragraph*{Fig~12.}
{\small\textbf{Notations used to characterize the position and motion of pedestrians:} (a) pedestrian $P_i$ is walking in a circle centered at  $C$, with velocity $\vec{v}_i$,  with an angle $\theta_{{\rm w}_i}$ relative to the normal to the wall, and at a distance $r_{{\rm w}_i}$ from the wall. 
	$\vec{e}_{{\rm w}_i}$ is the unit radial vector, $\vec{e}_{\parallel}=\vec{v}_i/v_i$ is the unit vector along the direction of motion, and $\vec{e}_{\perp}$ is the unit vector perpendicular to $\vec{e}_{\parallel}$; 
	%(b) $\vec{e}_{ij}$ the unit vector in the direction $\overrightarrow{P_i P_j}$ ($P_j$ is another pedestrian), $r_{ij}$ the distance between $P_i$ and $P_j$ and $\Psi_{ij}$ the angle between $\vec{v}_i$ and $\vec{e}_{ij}$.}
	(b) $\vec{e}_{ij}$ is the unit vector along the direction $\vec{P_i P_j}$  ($P_j$ is another pedestrian), $r_{ij}$ is the distance between $P_i$ and $P_j$, and $\Psi_{ij}$ is the (viewing) angle between $\vec{v}_i$ and $\vec{e}_{ij}$. }
\label{figangles}

\subsubsection*{Model for the random walk phase} 
The equation of motion for a given agent (model for a human subject) $i$ at position $\vec{x}_i$ and with velocity $\vec{v}_i$ generally reads  (see  \nameref{figangles} for a visual description of all  variables)
\begin{eqnarray} 
\frac{d \vec{v}_i}{dt} & =& \vec{A}(\vec{v}_i)  +   \sigma_0\,\vec{\eta}_i
+ \vec{F}_{{\rm w}_i} + \sum_{j=1, \, j \ne i}^N \vec{F}_{{\rm h}_{ij}}, \label{eqmain}\\
\frac{d\vec{x}_i }{ dt} & =& \vec{v}_i.
\end{eqnarray}
The first term in Eq~(\ref{eqmain}) represents a ``friction force'' which can be represented by the simple form (see  \nameref{fw+gw})
\begin{equation} 
\vec{A}(\vec{v}) =A(v) \, \vec{e}_{\parallel}= -\frac{v - v_0}{\tau_0} \, \vec{e}_{\parallel}\label{friction}.
\end{equation}
This friction term tends to make the modulus $v$ of the velocity close to the target speed $v_0$ and  is directed along the velocity of the agent, of direction $\vec{e}_{\parallel}=\vec{v}/v$.

The second term is a stochastic contribution that models the spontaneous fluctuations in the velocity of a subject, of intensity $\sigma_0$. The $x$ and $y$ coordinates of $\vec{\eta}_i(t)$ are independent  Gaussian noises, delta-correlated in time, and independent for different agents, $\langle \vec{\eta}_i(t)\vec{\eta}_j(t')\rangle =\delta_{i,j}\,\delta(t-t') $.

The third term is the repulsive interaction between the agent and the  boundaries of the circular arena and is of the form  (see  \nameref{fw+gw} and \cite{fishint})
\begin{equation}
\vec{F}_{\rm w} (v,r_{\rm w},\theta_{\rm w}) = - f_{\rm w}(r_{\rm w}) g_{\rm w}(\theta_{\rm w}) \vec{e}_{\rm w},\label{intwallprod}
\end{equation}
where $r_{\rm w}$ is the distance of the agent to the boundary/wall,  $\theta_{\rm w}$ is the relative angle between the velocity of the agent and the normal to the boundary, and $\vec{e}_{\rm w}$ is the unit radial vector between the center of the arena and the considered agent (and hence, anti-parallel to the normal to the boundary at its  closest point to the agent; see  \nameref{figangles} for a visual description of all these variables).
The interaction functions $f_{\rm w}(r_{\rm w})$ and $g_{\rm w}(\theta_{\rm w})$ represent the modulation of this repulsive interaction with $r_{\rm w}$ and $\theta_{\rm w}$, respectively. In particular, $g_{\rm w}(\theta_{\rm w})$ translates the human anisotropic perception of the environment (stronger response if the obstacle is ahead than behind). Both functions can be adequately represented by simple analytical forms:
\begin{eqnarray}
f_{\rm w}(r_{\rm w}) & =&
\left\{
\begin{array}{ll}
a_{\rm w} \left( e^{-\frac{r_{\rm w}}{l_{\rm w}}} - e^{-\frac{r_{\rm w_c}}{l_{\rm w}}} \right)  & \mbox{if } r_{\rm w} < r_{\rm w_c} \\
0 & \mbox{otherwise}
\end{array}
\right. \label{interfw}\\
g_{\rm w}(\theta_{\rm w}) & = & a_{\rm w_0} + a_{\rm w_1} \cos(\theta_{\rm w}) + a_{\rm w_4} \cos(4\,\theta_{\rm w}).\label{intergw}
%g_{\rm w}(\theta_{\rm w}) & =& \sum_{n=0}^4a_{{\rm w}_n}\cos(n\,\theta_{\rm w}) 
\end{eqnarray}
$r_{\rm w_c}$ is the critical distance to the wall beyond which pedestrians do not feel the effect of the wall anymore,  $l_{\rm w}$ is the  interaction range (values given in  \nameref{Table2}), and $a_{\rm w}$ is the intensity of the repulsion to the wall. $a_{{\rm w}_n}$ $(n=0,1,4)$ are the non-zero Fourier coefficients of $g_{\rm w}$ (see  \nameref{Table2}), which produce a function $g_{\rm w}(\theta_{\rm w}) $ sharply peaked for $\theta_{\rm w}$ between $-60^\circ$ and $+60^\circ$, as expected for human subjects (see \nameref{fw+gw}). Since multiplying $f_{\rm w}$ by a constant and $g_{\rm w}$ by the inverse of the same constant leaves the product in Eq~(\ref{intwallprod}) unchanged, we have normalized $g_{\rm w}$ so that its squared average is equal to 1:
\begin{equation}
\frac{1}{2\pi}\int_{-\pi}^\pi g_{\rm w}^2(\theta)\,d\theta=1.
\end{equation}
Note again that the forms of these functions were directly extracted from the data using a procedure analogous to the one described in \cite{fishint}, and hence, are not  mere assumptions.

\paragraph*{Fig~13.}\label{fw+gw}
{\small\textbf{Interaction functions characterizing human random walk}.
	%Dots are measures extracted from the data, while solid red lines are fitting analytical functions.
	Shape of
	(a) the self-propulsion force $A(v)$, 
	(b) the intensity of the interaction with the wall $f_{\rm w}(r_{\rm w})$ and with other pedestrians $f_{\rm h}(r_{ij})$, and
	(c) the modulation of the interaction intensity with the angle of incidence to the wall $g_{\rm w}(\theta_{\rm w})$. The modulation of the repulsive interaction $g_{\rm h}(\psi)$ between two subjects as a function of the viewing angle $\psi$ takes the same form as in (c).}\vskip 0.4cm

Finally, the last term in Eq~(\ref{eqmain}) is the sum of  the repulsive interactions between the agent $i$ and  all other agents. In general, the influence of agent  $j$ on agent $i$ depends on the distance $r_{ij} = |\vec{x}_i - \vec{x}_j|$ between them, on their relative speed $\vec{v}_{ij} = \vec{v}_i - \vec{v}_j$ (and in particular, on their relative orientation \cite{fishint}), and on the viewing angle $\psi_{ij}$ with which $i$ perceives~$j$ (i.e., the angle between the velocity of $i$ and the unit vector $\vec{e}_{ij}$ along the direction $\vec{i\,j}$; see  \nameref{figangles} for a visual description of all these variables)
\begin{eqnarray}
\vec{F}_{\rm h_{ij}} (\vec{v}_{ij},r_{ij},\psi_{ij}) = - B_{\rm h}(\vec{v}_{ij}) f_{\rm h}(r_{ij}) g_{\rm h}(\psi_{ij}) \vec{e}_{ij}.\label{interfhdef}
\end{eqnarray}
In the present work, we consider a simplified  model where $g_{\rm h}$  have the same functional form as $g_{\rm w}$ (see Eq~(\ref{intergw})), both reflecting the stronger reaction to an obstacle ahead than behind the agent:
\begin{equation}
g_{\rm h}(\psi_{ij})  = g_{\rm w}(\psi_{ij}).\label{intergh}
\end{equation}
We also take $B_{\rm h}(\vec{v}_{ij}) = 1$, and assume the following form for $f_{\rm h}(r_{ij})$:
\begin{equation}
f_{\rm h}(r_{ij}) =
\left\{
\begin{array}{ll}
a_{\rm h} \left( e^{- \left( \frac{r_{ij} }{ l_{\rm h}} \right)^2} - e^{- \left( \frac{r_{\rm h_c} }{ l_{\rm h}} \right)^2} \right)  & \mbox{if } r_{ij} < r_{\rm h_c} \\
0 & \mbox{otherwise}
\end{array}
\right.\label{interfh}
\end{equation}
where $r_{\rm h_c}$ is the critical distance between individuals, beyond which the interaction vanishes, and $l_{\rm h}$ is the typical interaction range between humans (see parameter values in   \nameref{Table2}).

Note that the viewing angles are in general not symmetric ($\psi_{ij}\ne \psi_{ji}$) so that the ``force'' exerted by $j$ on $i$ is not simply the opposite of the one exerted by $i$ on $j$. This general  breaking of the usual Newtonian law of action-reaction for social interactions can have important consequences for the collective dynamics of human and animal groups \cite{fishint}.

The results of the model shown in \nameref{fig:2} for the random walk phase (lasting typically 45\,s in each experimental run before the segregation phase) were produced by discretizing the equation of motion Eq~(\ref{eqmain}) in time with a typical time step $dt$ less than $0.01$\,s and conducting 1000 simulations of 3 minutes effective duration each, so that the error bars for the results of the model are negligible on the scale of the different graphs presented in the present work.

\subsubsection*{Model for the segregation phase} 
In order to describe the segregation phase, we adopt the same model as above when the agent is beeping (and is hence assumed to move). However, we found  that the actual subjects were walking more slowly than in the random walk phase, as they were paying attention to the sound emitted by the tag clipped on their left shoulder. Hence, the parameter $v_0$  (as well as some other parameters) is replaced by a smaller target speed $v_{\rm beep}$ and the relaxation time $\tau_0$ is replaced by $\tau_{\rm beep}$ in  Eq~(\ref{friction}) defining the friction term $A(v)$ (see parameter values in   \nameref{Table3}):
\begin{equation}
A(v)  = -\frac{v - v_{\rm beep}}{\tau_{\rm beep}}.\label{eqfin0}
\end{equation}

When an agent is not beeping, we assume a simple strategy where the agent decides to stop, being satisfied with its current position. Hence, its new target velocity is zero in Eq~(\ref{friction}), the friction term becoming
\begin{equation}
A(v)=-\frac{v}{\tau},\label{eqfin}
\end{equation}
(with the value of $\tau$ given in   \nameref{Table3}), and the noise intensity reflecting the spontaneous fluctuations of its velocity also vanishes, while the interaction with the arena boundary and the other agents is preserved.

\subsection*{Data availability}

The data  supporting the findings of this study are available at \textit{fig\textbf{share}}:\\
\noindent\href{https://doi.org/10.6084/m9.figshare.9897590}{https://doi.org/10.6084/m9.figshare.9897590}.

\section*{Acknowledgements}
We thank Mathieu Moreau and G\'erard Latil for their help in preparing and running the experiments. This work was supported by Agence Nationale de la Recherche project 11-IDEX-0002-02–Transversalit\'e–Multi-Disciplinary Study of Emergence Phenomena, a grant from the CNRS Mission for Interdisciplinarity (project SmartCrowd, AMI S2C3), and by Program Investissements d’Avenir under Agence Nationale de la Recherche program 11-IDEX-0002-02, reference ANR-10-LABX-0037-NEXT. B.\,J. was supported by a doctoral fellowship from the CNRS, and R.\,E. was supported by Marie Curie Core/Program Grant Funding Grant 655235–SmartMass. The funders had no role in study design, data collection and analysis, decision to publish, or preparation of the manuscript.

\section*{Author contributions} M.\,R., C.\,S., G.\,Th., and G.\,Tr.~designed the experiments. All authors conducted the experiments. M.\,R., R.\,P., G.\,Tr., and C.\,Z. were responsible for the data collection. R.\,E. and C.\,S. performed the data treatment. B.\,J.  conducted the statistical analysis with input from R.E., C.S., and G.\,Th. The model was designed by C.\,S., with input from R.\,E., B.\,J., and G.\,Th. Numerical simulations of the model were performed by R.\,E. and B.\,J, with inputs from C.\,S. and G.\,Th. Finally, R.\,E., B.\,J., C.\,S. and G.\,Th. wrote the article and all authors contributed to critically revising the manuscript.

\newpage
\section*{Figures}

\begin{figure*}[hpb]
	\centering
	\includegraphics[width=0.98\textwidth]{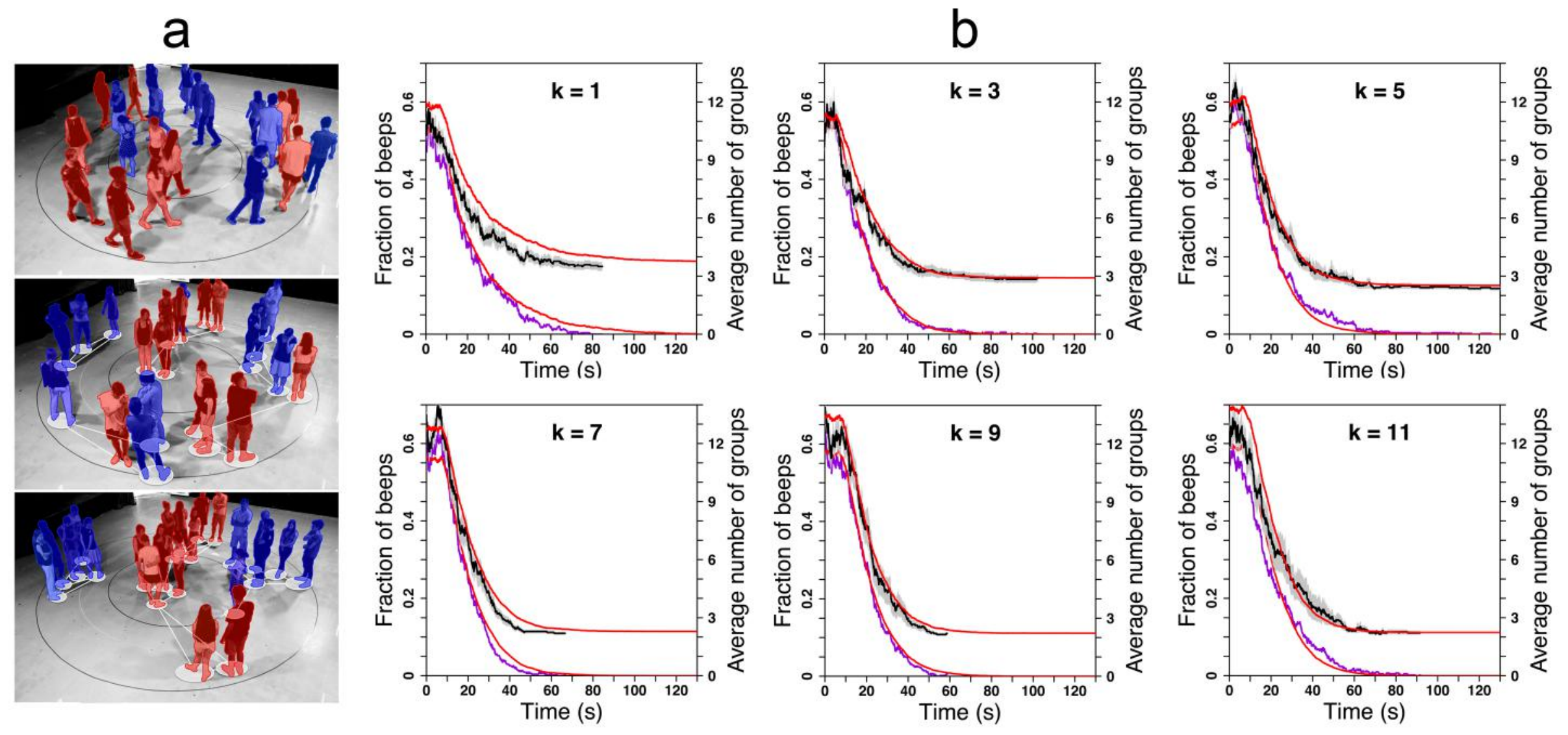}
\end{figure*}
\paragraph*{Fig~1.}
{\label{fig:1}\textbf{Illustration of the different stages of an experimental run.} (\textbf{a})~Random walk phase for typically 45 seconds (top panel), phase separation process (beeping phase; middle panel), and final silent state where the subjects do not beep and do not move anymore (bottom panel), exemplified for an experiment with $k = 5$.
	(\textbf{b})~Decay of the fraction of beeping subjects (purple lines) and of the number of 3-groups (black lines) averaged over all experimental runs for $k=1$, 3, 5, 7, 9, and 11, compared to the results of the model (red smooth lines).}
\FloatBarrier

\newpage
\begin{figure*}[htp]
	\centering	
	\includegraphics[width=0.95 \textwidth]{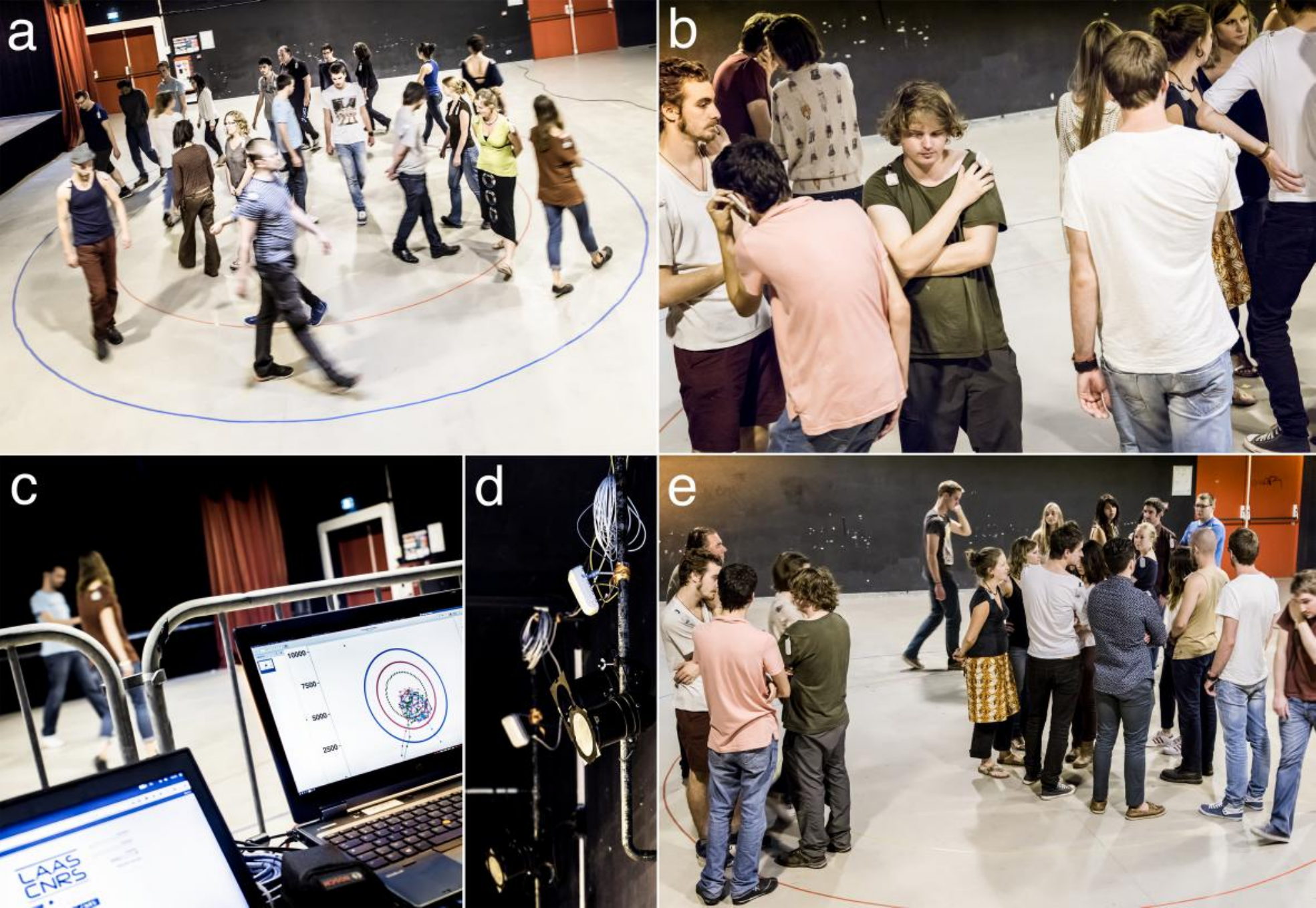}
\end{figure*}
\paragraph*{Fig~2.}
{\textbf{Segregation experiment and tracking system}. An experimental run starts with the random
	color assignment to the 22~participants (11~blue and 11~red) and the random walking
	phase (a), where participants walk in a circular arena delimited with a blue tape on the floor.
	In the segregation phase (b, c, e), the tag clipped on each participant left shoulder emits
	a beep if the color of the majority of his/her $k$ nearest neighbors ($k = 1$, 3, 5, 7, 9, 11, and 13) is different from 	his/her own. The positions of the participants are  recorded in real time by the Ubisense tracking system (c, d) based on the triangulation of sensors (d) located around the	arena.}
\label{setup}
\FloatBarrier

\newpage
\begin{figure*}[h!]
	\centering		
	\includegraphics[width=0.95 \textwidth]{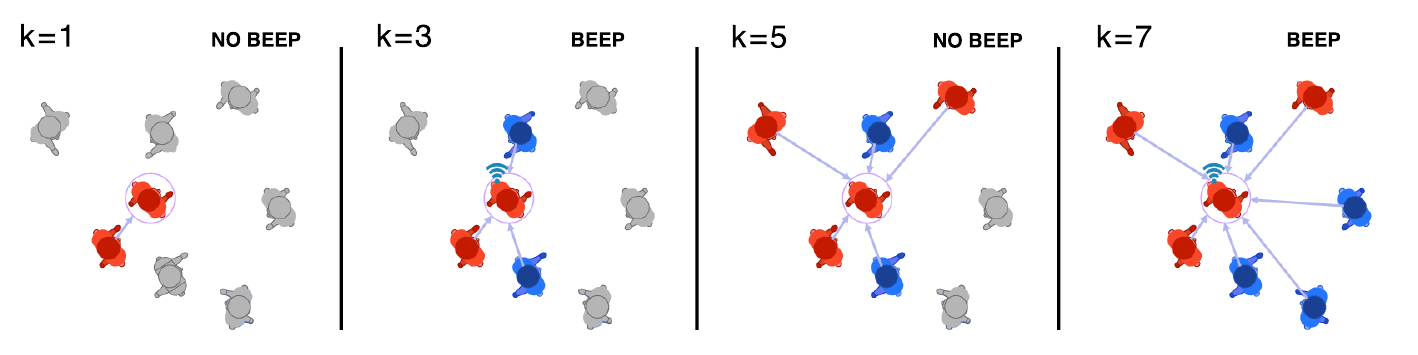}
\end{figure*}
\paragraph*{Fig~3.}
{\textbf{Illustration of the beeping state}:
	the red focal individual is marked by a circle, and is connected to its $k$ nearest neighbors by arrows. 
	The tags on his/her left shoulder beeps when the majority of his/her $k$ nearest neighbors is of a different color from him/her (for $k = 3$ and 7 above, as illustrated by the blue wave-like symbol). 
	Grey neighbors are beyond the interaction range, and are thus not considered in the ‘‘environment’’ of the focal individual (i.e., in the computation of the beep).}
\label{beepingcases}
\FloatBarrier

\newpage
\begin{figure*}[ht]
	\centering		
	\includegraphics[width=0.95 \textwidth]{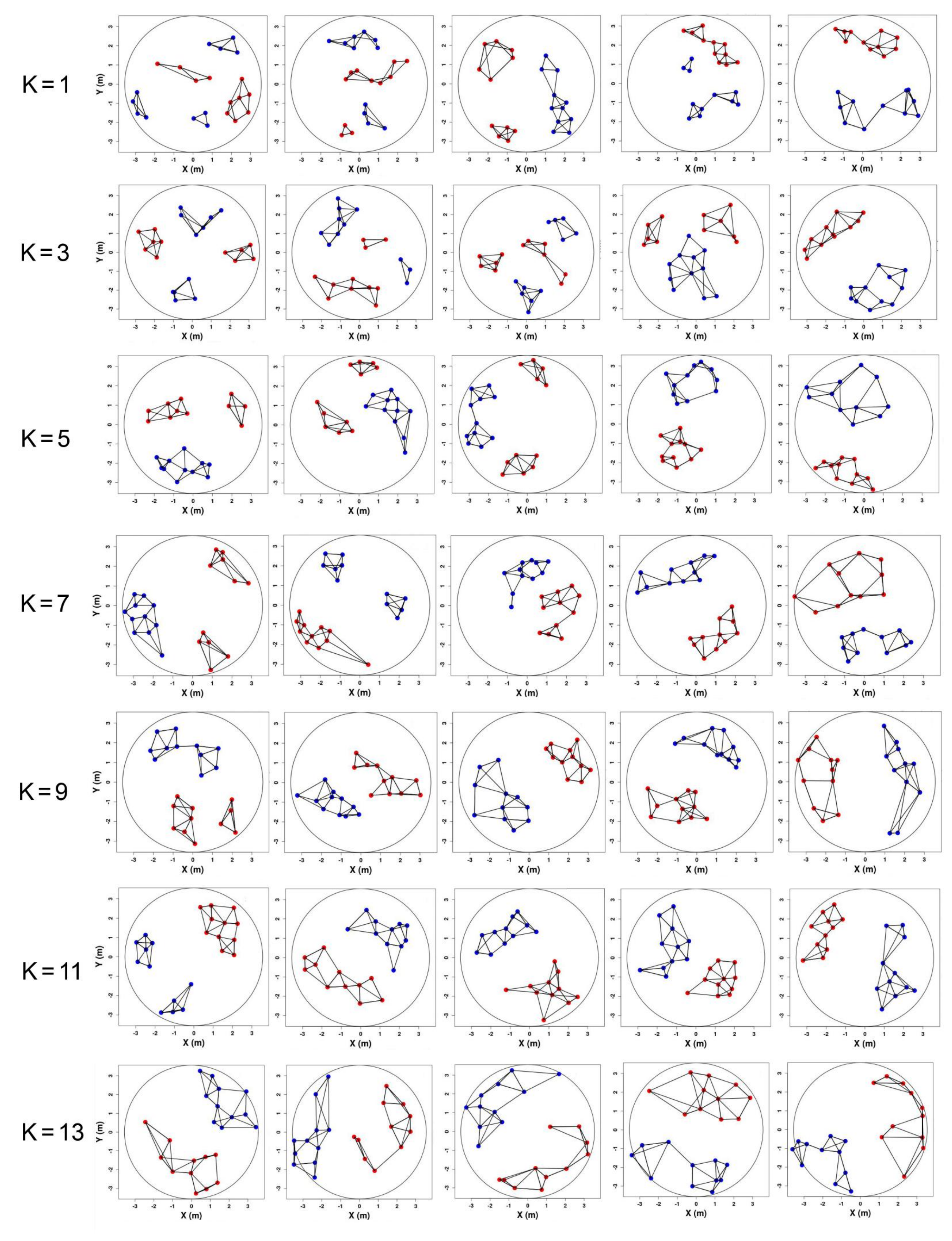}
\end{figure*}
\paragraph*{Fig~4}
{\textbf{Representative selection of final states for all values of ${k}$}. The quality of the separation increases (i.e. less groups are formed in the final state) with $k$ up to $k=7$, value above which it saturates.}
\label{final_config_1}
\FloatBarrier

\newpage
\begin{figure*} [h!]
	\centering
	\includegraphics[width=1 \textwidth]{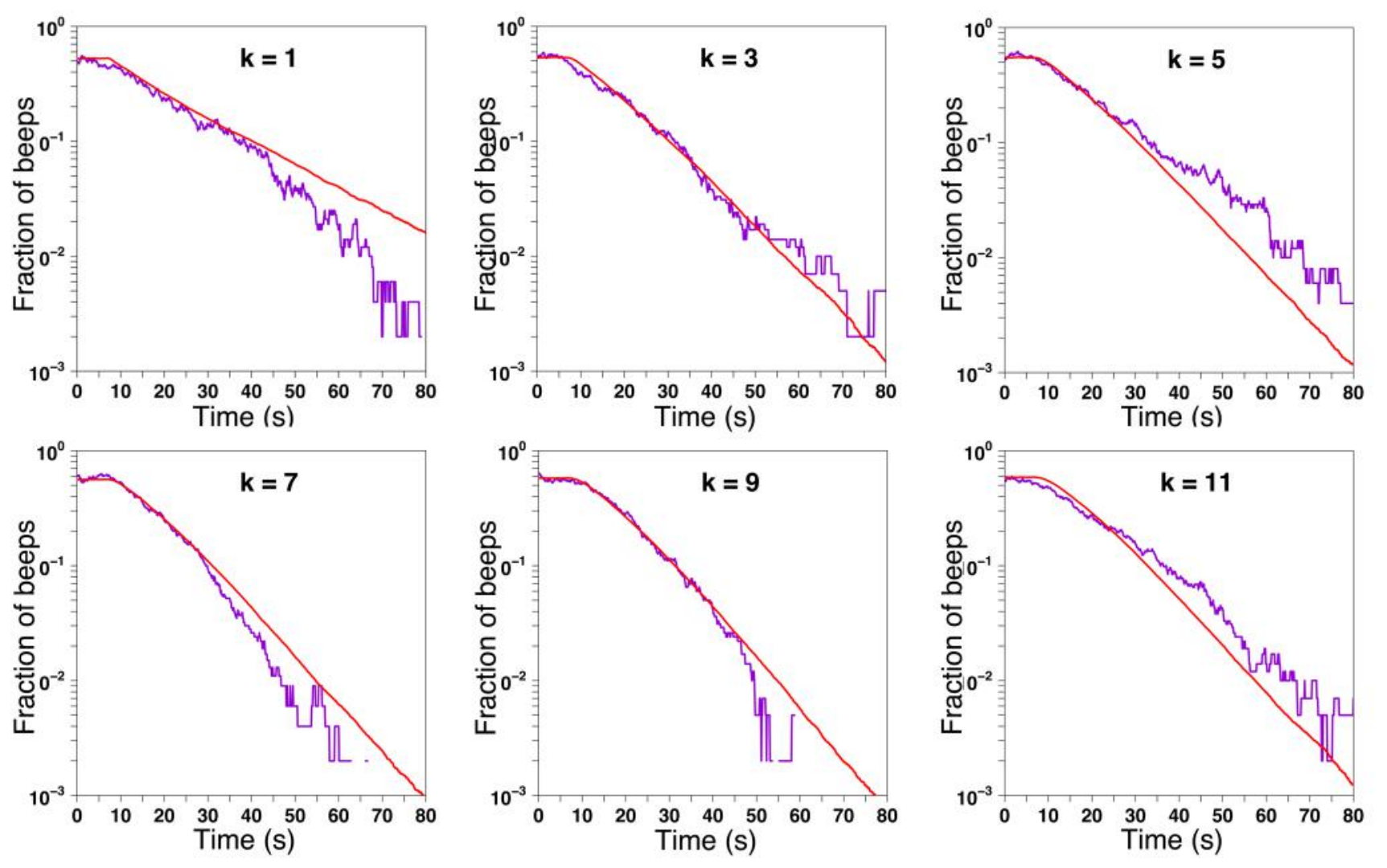}
\end{figure*}
\paragraph*{Fig~5.}
{\textbf{Instantaneous fraction of beeps for each value of ${k}$}:
	Purple lines show the experimental data and red lines model simulations.
	The y-axis is in log scale, evidencing the near-exponential decay.} \label{NumberGroupslog}
\FloatBarrier

\newpage
\begin{figure}[htp]
	\centering
	\includegraphics[width=0.98\linewidth]{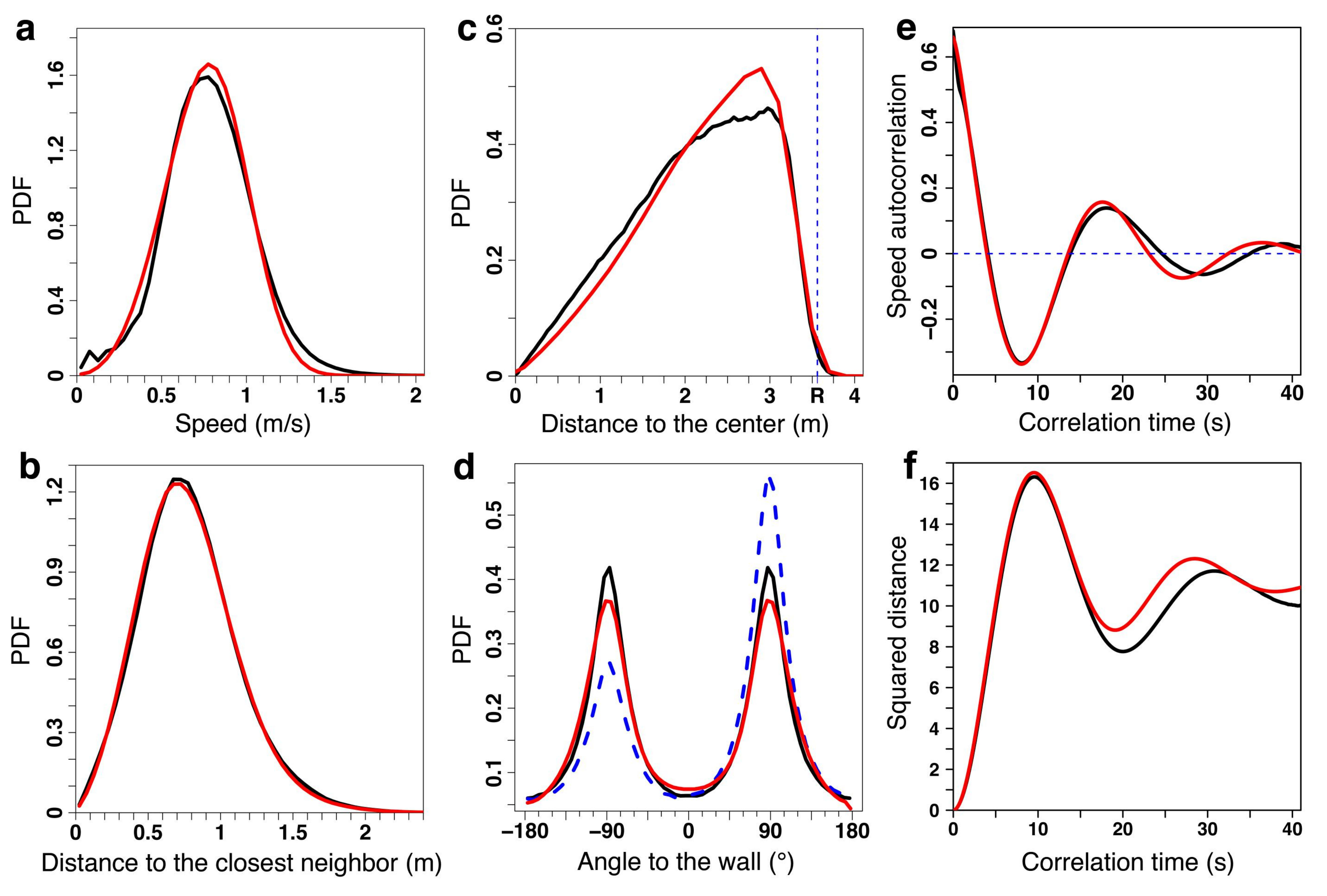}
\end{figure}
\paragraph*{Fig~6.}
{\label{fig:2}\textbf{Quantitative analysis of the model for 22 subjects in the random walk phase.} The experimental and  the model results are respectively plotted in black and red. (\textbf{a})~Probability distribution function (PDF) of the instantaneous speed of the subjects (agents in the model).   (\textbf{b})~PDF of the distance between a subject/agent and its closest neighbor at a given time. (\textbf{c})~PDF of the distance of the subjects/agents from the center of the arena. (\textbf{d})~PDF of the relative angle between the velocity of a subject/agent and the normal to the boundary of the circular arena. The raw data (blue dashed line) show that the subjects in the experiment turn more often anticlockwise, keeping the boundary of the arena to their right. The black curve is obtained by symmetrizing the data, i.e., by adding all mirror trajectories to the data set. The agent model does not have any intrinsic left/right asymmetry. (\textbf{e})~Time correlation function of the velocity $C_v(t)=\langle\vec{v}(t+t')\vec{v}(t')\rangle$, where $\langle\,{\rm \bullet}\,\rangle$ stands for the average over all experimental runs and over the reference time $t'$. Oscillations translate the velocity anti-correlation developing due to the bounded arena. (\textbf{f})~Mean square displacement of a subject/agent $D_x(t)=\langle[\vec{x}(t+t')-\vec{x}(t')]^2\rangle$. For large time, when the trajectories decorrelate, it ultimately saturates to $2\langle\vec{x}^2(t')\rangle$, i.e., twice the mean square distance between a subject/agent and the center of the arena (at $\vec{C}=(0,0))$, which can also be evaluated from (\textbf{c}).}
\FloatBarrier

\newpage
\begin{figure*}[h!]
	\centering		
	\includegraphics[width=0.95 \textwidth]{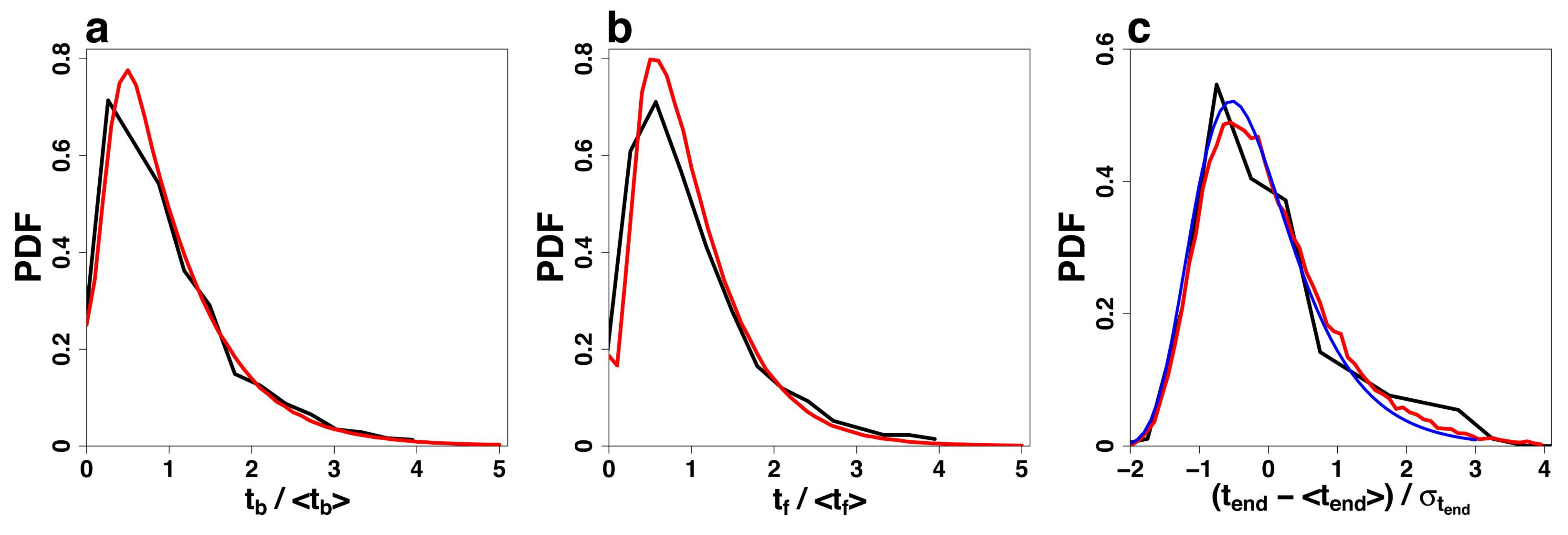}
\end{figure*}
\paragraph*{Fig~7.}
{\textbf{Probability density functions (PDF) of the normalized segregation time}.
	(a) $\bar{t}_{\rm b} = t_{\rm b}/\left< t_{\rm b}\right>$, where $t_{\rm b}$ is the total
	time a given individual spends beeping.
	(b) $\bar{t}_{\rm f} = t_{\rm f}/\left< t_{\rm f}\right>$, where $t_{\rm f}$ is the final
	time a given individual has beeped.
	(c) $\bar{t}_{\rm end} = (t_{\rm end} - \left< t_{\rm end}\right>)/\sigma_{t_{\rm end}}$,
	where $t_{\rm end}$ is the maximum value of the  $t_{\rm f}$ for a given experimental run, and hence coincides with the total duration of this run.
	Black lines: experimental data; red lines: model simulations. The blue line in (c) is the
	universal Gumbel distribution characterizing the distribution of the maximum of independent (or weakly dependent) random variables ($t_{\rm end}$ is the maximum of the individual ${t}_{\rm f}$).
	In (a) and (b), mean values $\left< \bullet \right>$ take into account all pedestrians from
	all sessions and all values for $k$; in (c), the mean value is calculated over all sessions
	and all values for $k$.}
\label{time-PDFs}
\FloatBarrier

\newpage
\paragraph*{Fig~8.}
\begin{figure}[t]
	\centering
	\includegraphics[width=0.98\linewidth]{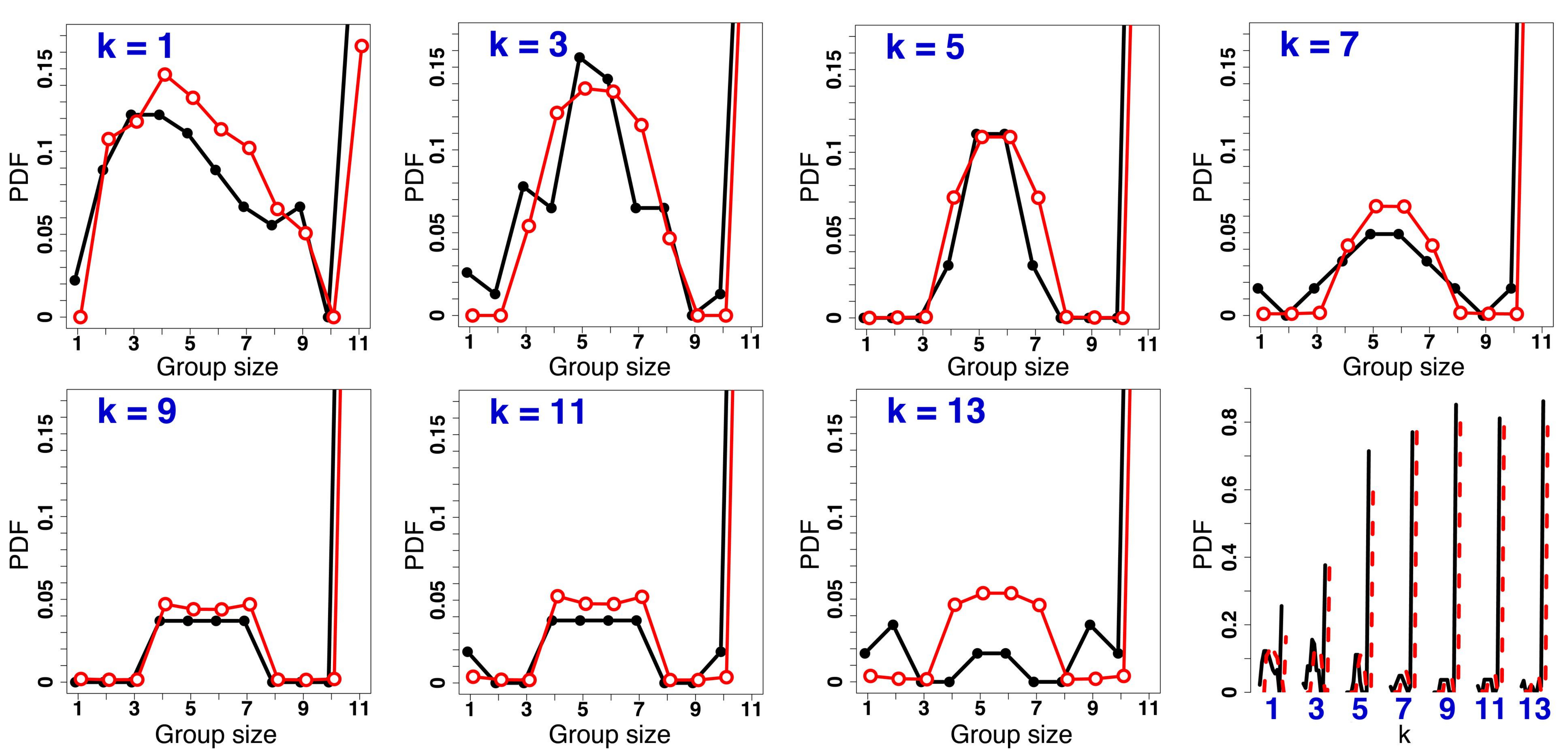}
\end{figure}
{\label{fig:3}\textbf{Characterization of the unicolor groups in the final silent state.} The experimental and  the model results are respectively plotted in black and red (the model plots are slightly shifted to the right for better readability). Probability distribution function (PDF) of the size of the 3-groups in the final silent state are shown for $k=1,3,5,7,9,11,13$. The vertical scale is adapted to focus on the probabilities to find a group of size strictly less than 11. Note that due to the experimental noise in the positions of the subjects, a few groups of size 1 and 2 were experimentally found in some instances. The last panel on the bottom right summarizes the results using a larger vertical scale to reveal the increase and the ultimate saturation of the fraction of groups of size $22/2=11$.}
\FloatBarrier

\newpage
\begin{figure}[htp]
	\centering
	\includegraphics[width=0.8\linewidth]{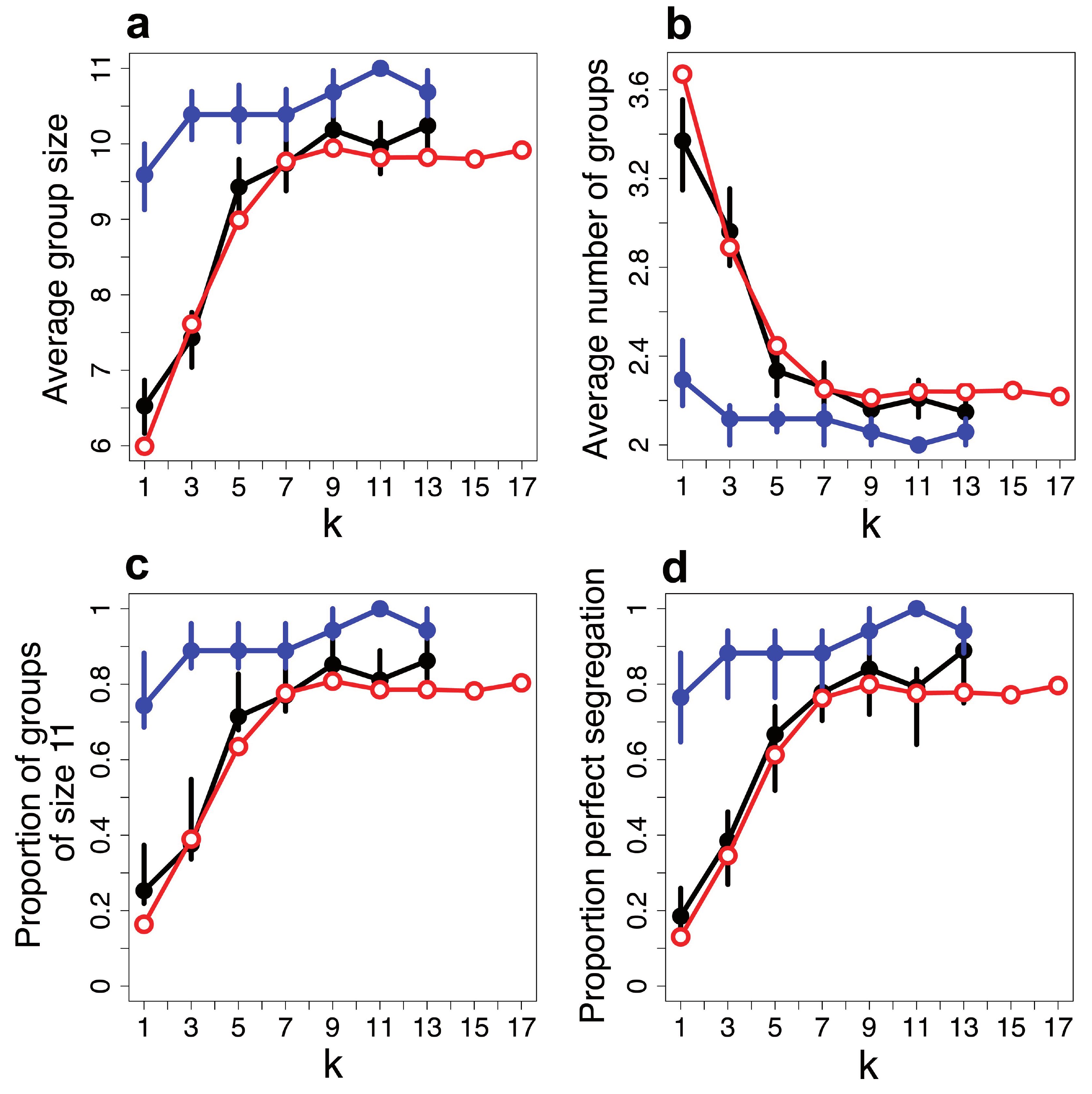}
\end{figure}
\paragraph*{Fig~9.}
{\label{fig:4}\textbf{Characterization of the efficiency of the separation as a function of $k$.} The experimental and  the model results are respectively plotted in black and red. The blue symbols and lines correspond to the results of a series of experiments where the subjects were additionally instructed to form  two clearly identifiable groups in the final state. (\textbf{a})~Average size of the 3-groups in the final silent state, as a function of $k$.  (\textbf{b})~Average number of groups. (\textbf{c})~Fraction of groups of maximum size 11. (\textbf{d})~Probability to observe a perfect separation in two groups of size 11. All panels illustrate the improvement of the segregation in the main experiment and the model up to $k=7$, above which it saturates. For the experiment with the additional instruction of forming two clearly identifiable groups, the efficiency of the segregation is much improved even for small $k$, and only slightly increases with $k$ (blue circles and line).}
\FloatBarrier

\newpage
\begin{figure*}[h!]
	\centering		
	\includegraphics[width=0.95 \textwidth]{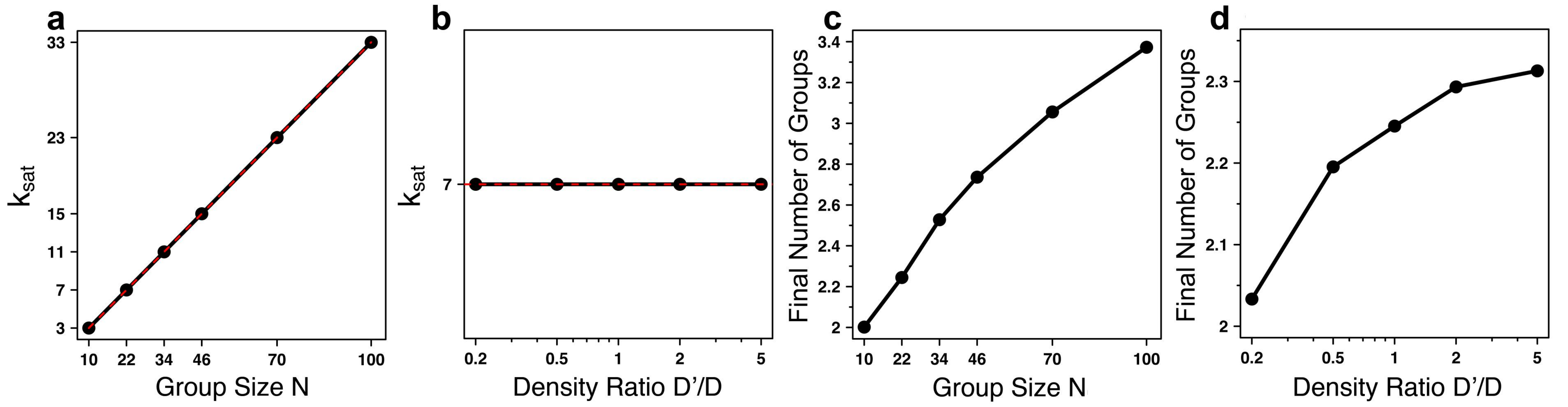}
\end{figure*}
\paragraph*{Fig~10.}
{\textbf{Saturation analysis for different group sizes and densities}: Values of $k$ above which the average number of groups -- and other related quantities (see \nameref{fig:4}) -- saturates, for different (a) group sizes $N$ and (b) pedestrian densities $D'$ with $N = 22$ agents (like in the experiment; $D$ is the actual experimental density of 0.55 pedestrian/m$^2$), and the corresponding saturation values of the average number of groups (c and d). $k_{\rm sat}$ is found to grow like $\approx N/3$ (red dashed line in a) and to be essentially independent of the density (red dashed line in b).}
\label{SaturationSI}
\FloatBarrier

\newpage
\begin{figure*}[ht]
	\centering		
	\includegraphics[width=0.95 \textwidth]{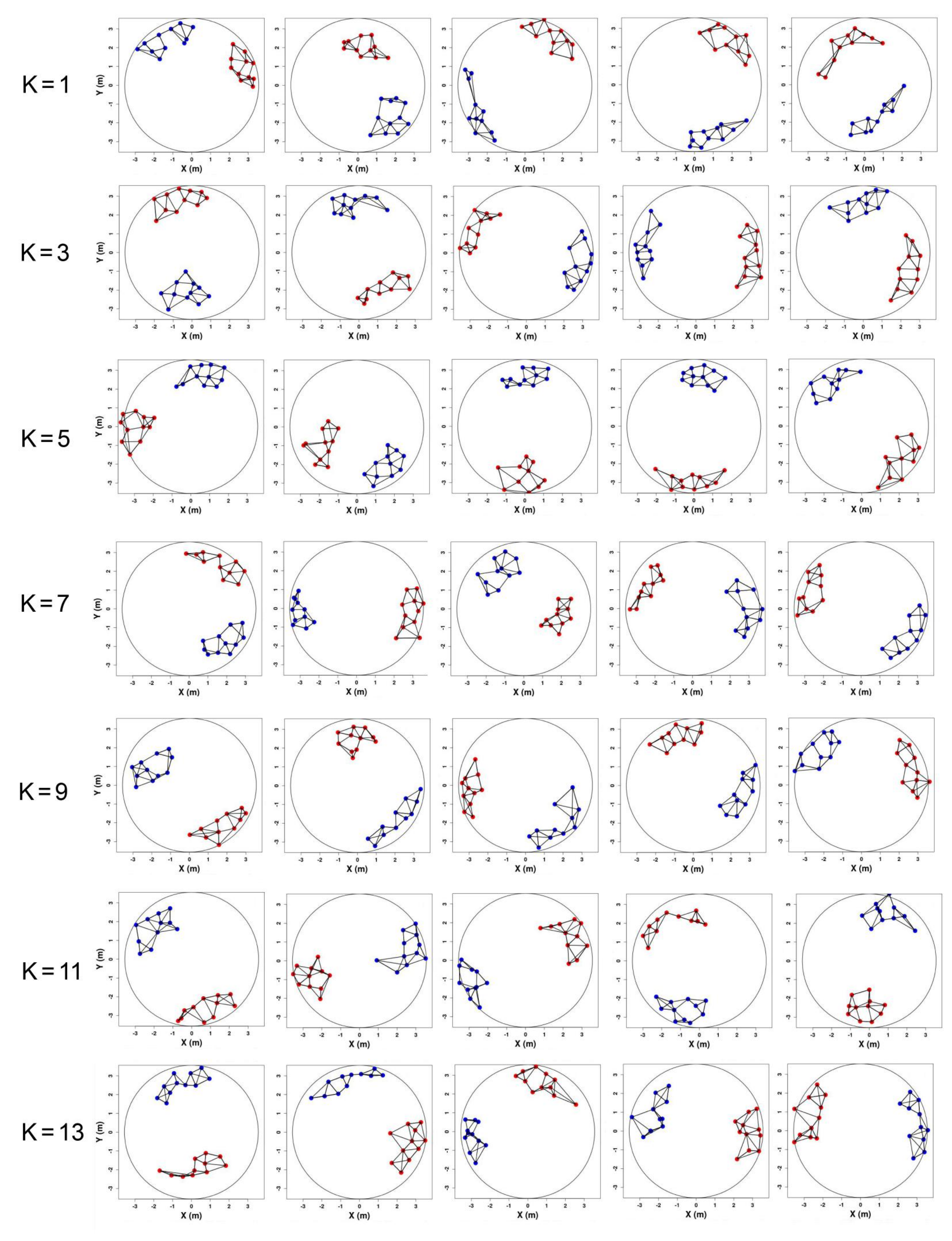}
\end{figure*}
\paragraph*{Fig~11.}
{\textbf{Representative selection of final states for all values of $k$}, in the experiments where subjects were explicitly asked to separate in two distinct groups.
	The separation was then perfect in most cases.}
\label{final_config_2}
\FloatBarrier

\newpage
\begin{figure*} [t]
	\centering
	\includegraphics[width=0.85 \textwidth]{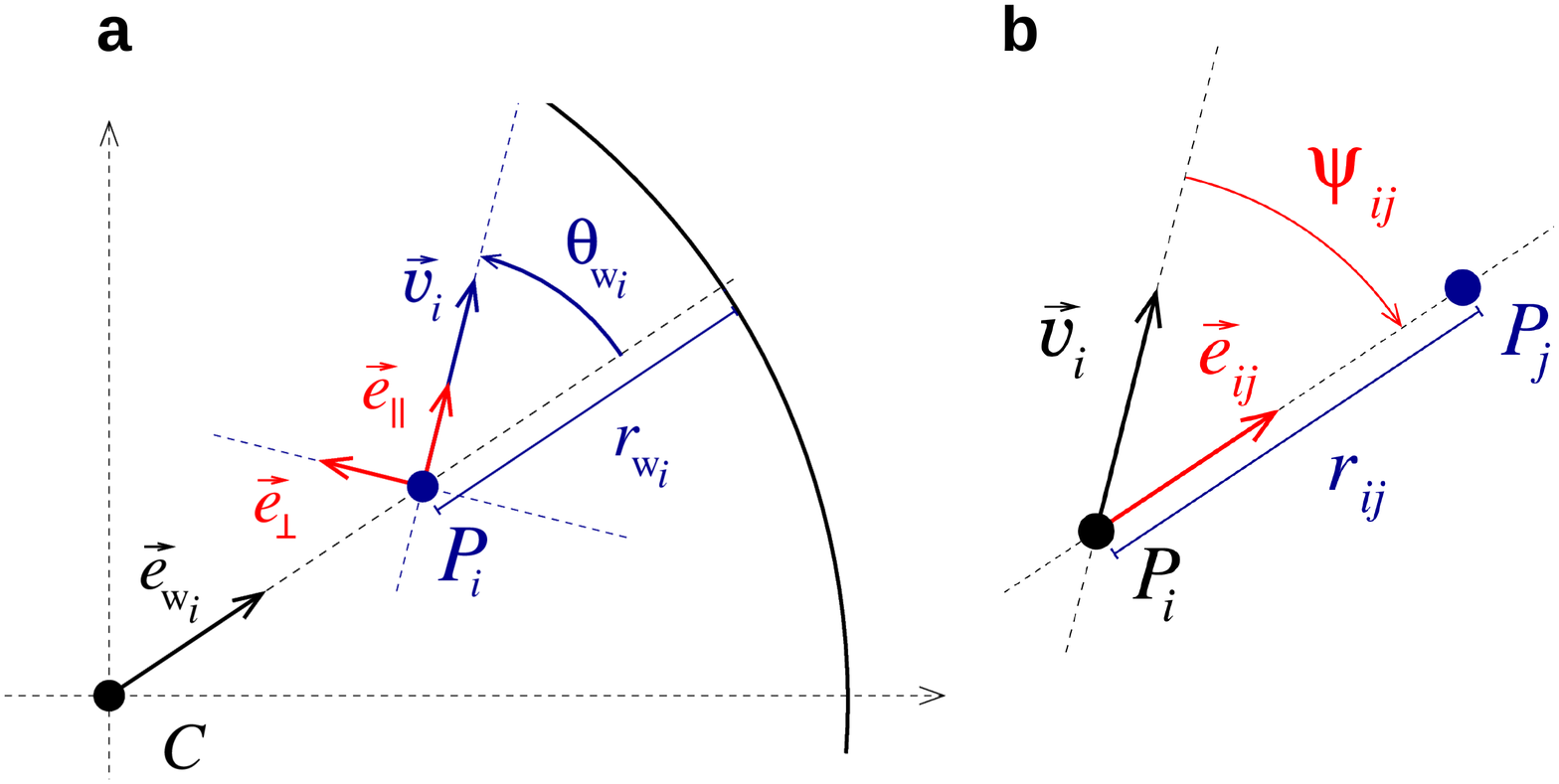}
\end{figure*}
\paragraph*{Fig~12.}
{\textbf{Notations used to characterize the position and motion of pedestrians:} (a) pedestrian $P_i$ is walking in a circle centered at  $C$, with velocity $\vec{v}_i$,  with an angle $\theta_{{\rm w}_i}$ relative to the normal to the wall, and at a distance $r_{{\rm w}_i}$ from the wall. 
	$\vec{e}_{{\rm w}_i}$ is the unit radial vector, $\vec{e}_{\parallel}=\vec{v}_i/v_i$ is the unit vector along the direction of motion, and $\vec{e}_{\perp}$ is the unit vector perpendicular to $\vec{e}_{\parallel}$; 
	%(b) $\vec{e}_{ij}$ the unit vector in the direction $\overrightarrow{P_i P_j}$ ($P_j$ is another pedestrian), $r_{ij}$ the distance between $P_i$ and $P_j$ and $\Psi_{ij}$ the angle between $\vec{v}_i$ and $\vec{e}_{ij}$.}
	(b) $\vec{e}_{ij}$ is the unit vector along the direction $\vec{P_i P_j}$  ($P_j$ is another pedestrian), $r_{ij}$ is the distance between $P_i$ and $P_j$, and $\Psi_{ij}$ is the (viewing) angle between $\vec{v}_i$ and $\vec{e}_{ij}$. }
\label{figangles}
\FloatBarrier

\newpage
\begin{figure*}[!htbp]
	\centering		
	\includegraphics[width=0.95 \textwidth]{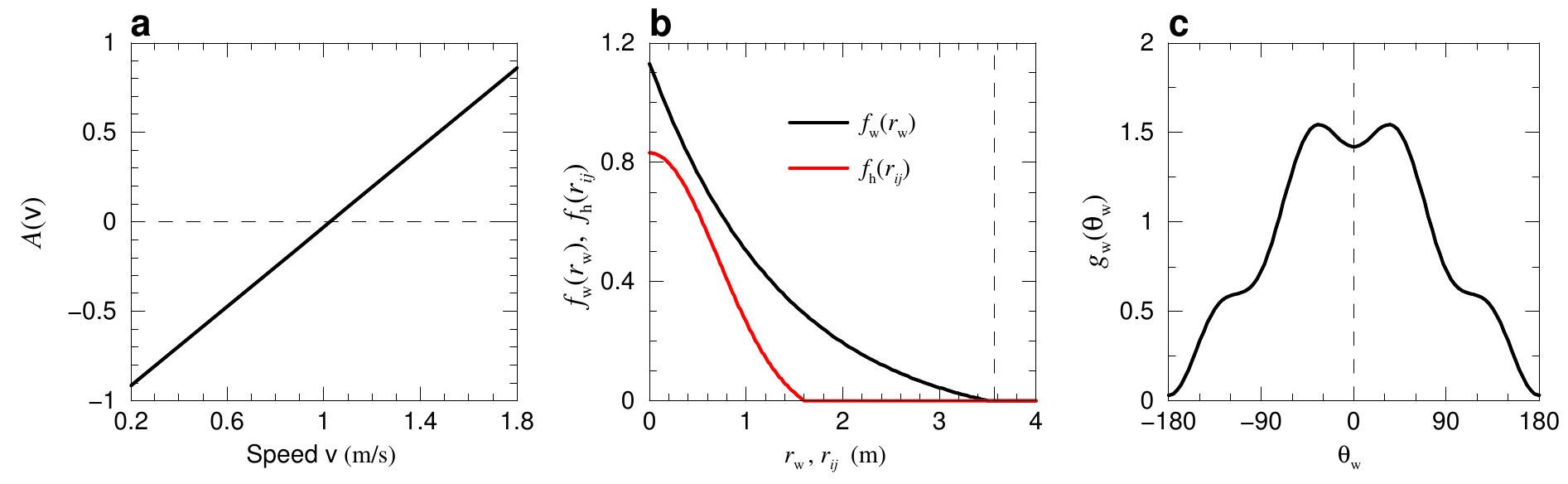}
\end{figure*}
\paragraph*{Fig~13.}\label{fw+gw}
{\textbf{Interaction functions characterizing human random walk}.
	Shape of
	(a) the self-propulsion force $A(v)$, 
	(b) the intensity of the interaction with the wall $f_{\rm w}(r_{\rm w})$ and with other pedestrians $f_{\rm h}(r_{ij})$ and
	(c) the modulation of the interaction intensity with the angle of incidence to the wall $g_{\rm w}(\theta_{\rm w})$. The modulation of the repulsive interaction between two subjects as a function of the viewing angle $\psi$ takes the same form as in (c).
}
\FloatBarrier

\newpage
\section*{Supporting Information}

\subsection*{Supplementary figures}

\vskip 0.4cm
\begin{figure*}[h!]
	\centering		
	\includegraphics[width=0.95 \textwidth]{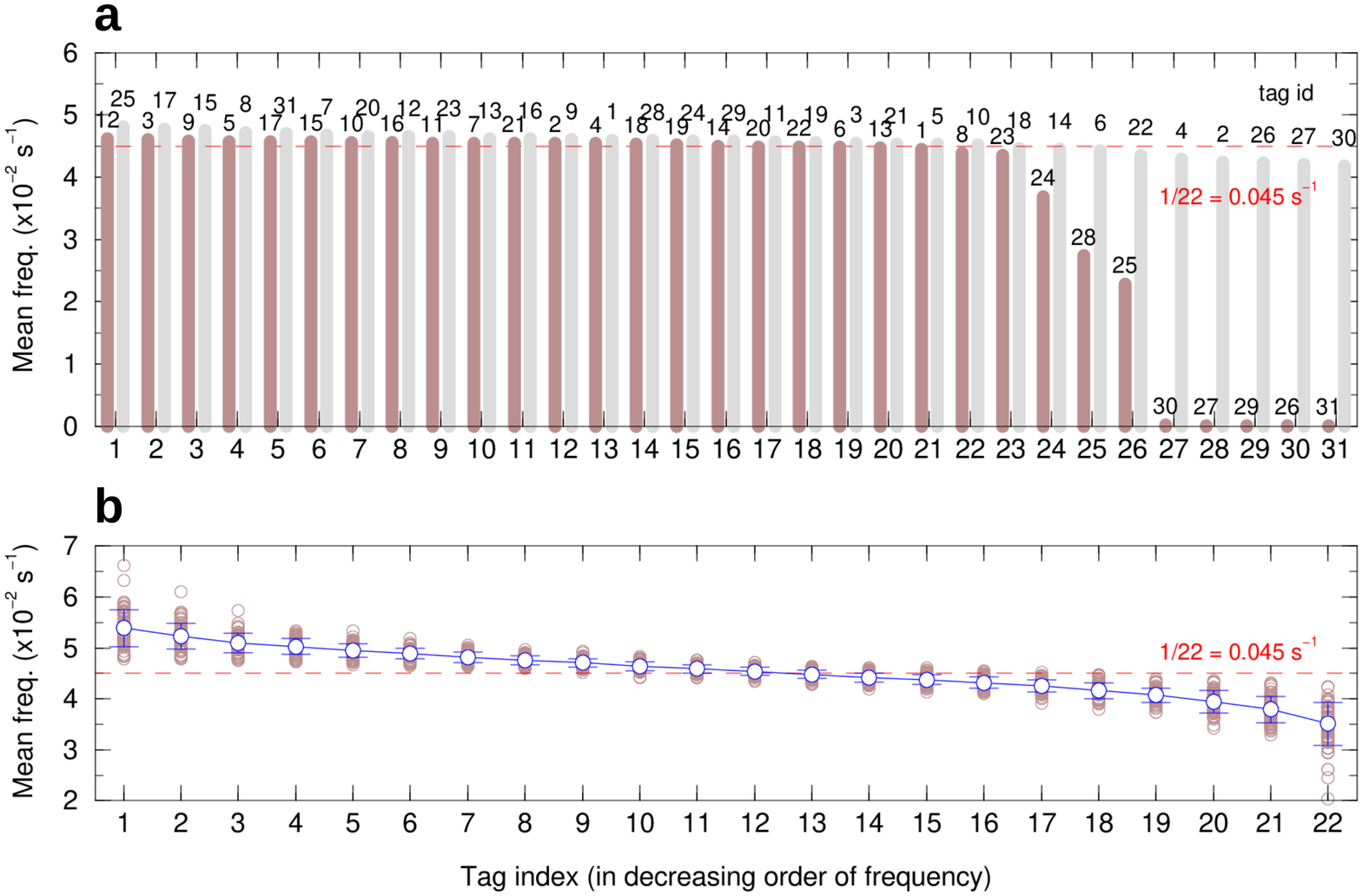}
\end{figure*}
\paragraph*{S1~Fig.}
{\textbf{Mean frequency of the number of data sent by a tag during a session}.
	(a) Mean frequency of each tag, indicated by its identity number. 
	Brown data correspond to September 2015, gray data to June 2016, where more than 30
	tags were used during the experiments (due to replacement of batteries or individual).
	(b) Mean frequency of a session's most frequent tag, second most
	frequent tag, and so on, until the session's least frequent tag. The most frequent
	tag is not always the same from one session to another.
	The theoretical value is $1/22=0.045$~s$^{-1}$ (red dashed lines).}
\label{tag-freq}
\FloatBarrier

\newpage
\begin{figure*}[h!]
	\centering		
	\includegraphics[width=0.95 \textwidth]{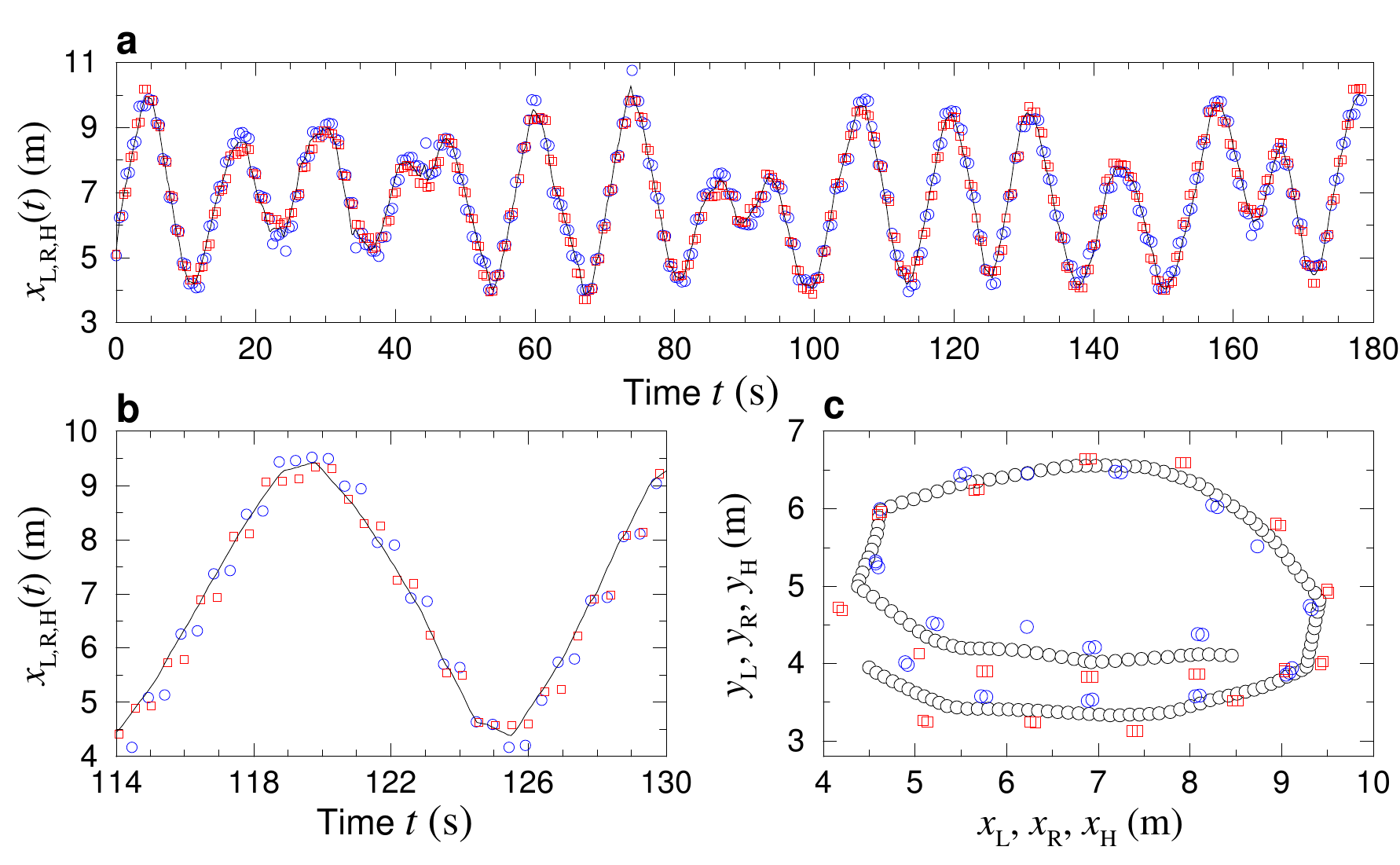}
\end{figure*}
\paragraph*{S2~Fig.}
{\textbf{Trajectory Reconstruction}.
	(a) Time series $x_{\rm L}(t)$ (blue circles) and $x_{\rm R}(t)$ (red squares) of the $x$-coordinates of an individual's left and right tags respectively,
	as directly extracted from the data collected. The positions are not uniformly distributed in time, with condensed packs of data (``bursts'') at some time positions, and sparse data at others.
	(b) Detail of the time interval $[114.5, 128.5]$, showing that tags' time series
	alternate (square--circle), while spatial series of each tag appear by pairs
	(two squares--two circles), with an alternating spatial gap of size $\approx 0.05$~m
	between spatially consecutive data and of size $\approx 1$~m between alternating pairs.
	The black lines in (a) and (b) show the synchronized time series (interpolated with
	constant time-step $\Delta t = 0.1$~s) of the $x$-coordinate $x_{\rm H}(s)$ of the geometric center of the two tags;
	(c) Final reconstructed trajectory (\emph{i.e.} after rectification of the ``burst'' effect; black circles) with the same time interval $\Delta s = \Delta t = 0.1$~s
	as in (b), and successive (real) positions of both tags.	}
\label{traj-reconstr}
\FloatBarrier

\newpage
\begin{figure*}[h!]
	\centering		
	\includegraphics[width=0.95 \textwidth]{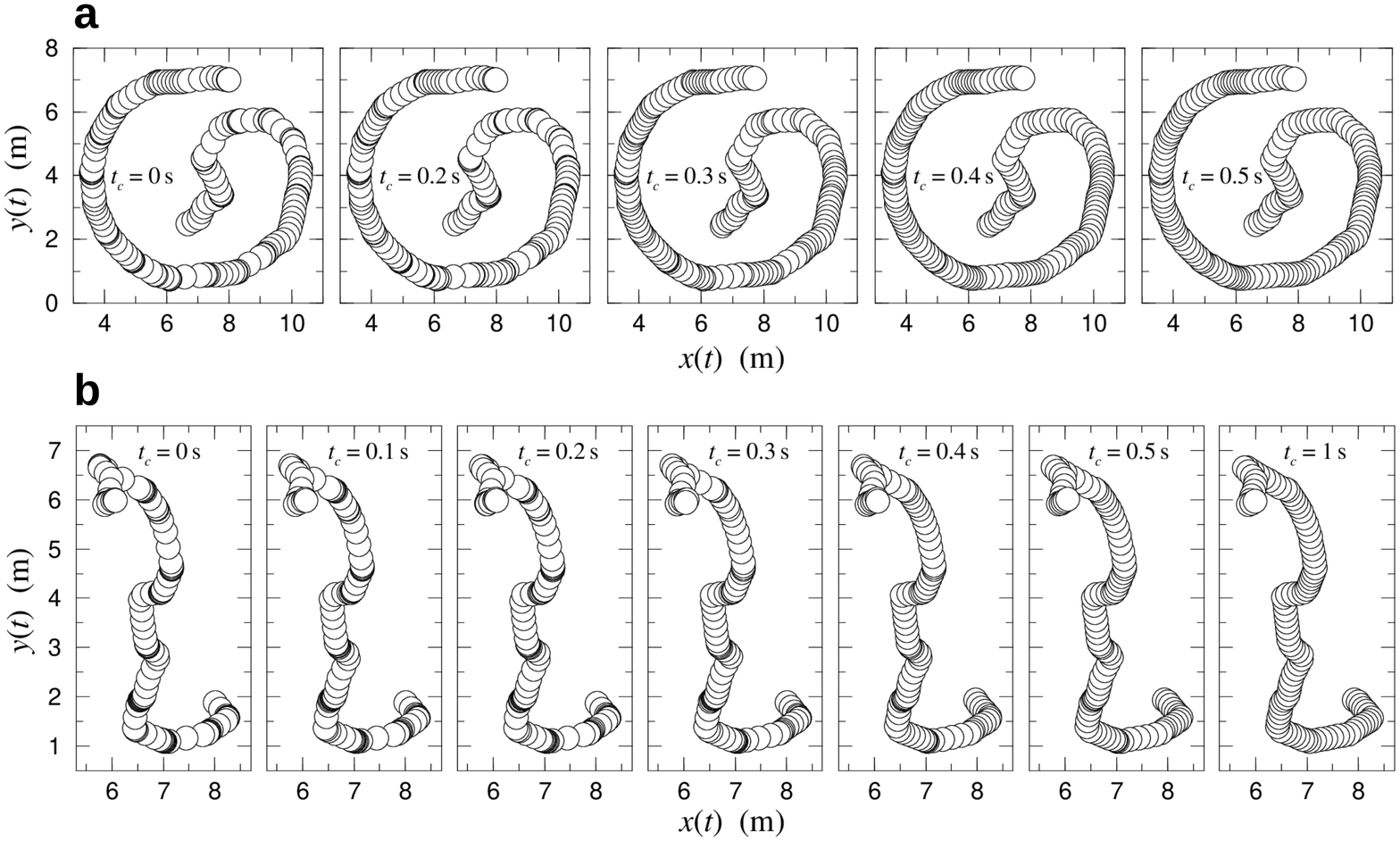}
\end{figure*}
\paragraph*{S3~Fig.}
{\textbf{Example of individual trajectory}.
	%\caption{\doublespacing\textbf{Individual trajectory}:
	(a) Trajectory of an individual during the random walk phase.
	(b) Trajectory of the same individual during the segregation phase.
	Circles denote successive positions of the individual at equally spaced time instants
	($\Delta t = 0.2$\,s). Black rings observed for small values of $t_c$ correspond to the
	accumulation of circles in short intervals of time, i.e., a ``burst''.
	The panels illustrate the reduction of the burst effect, for increasing values of the averaging parameter $t_c$, through the redistribution of time instants.
	$t_c=0.4$\,s (which we selected for the reconstruction of our data) offers a good compromise between small values for which artificial bursts are excessively pronounced, and large values for which real variations in speed are smoothed out.}
\label{datawaves}
\FloatBarrier

\newpage
\begin{figure*}[!htbp]
	\centering		
	\includegraphics[width=0.95 \textwidth]{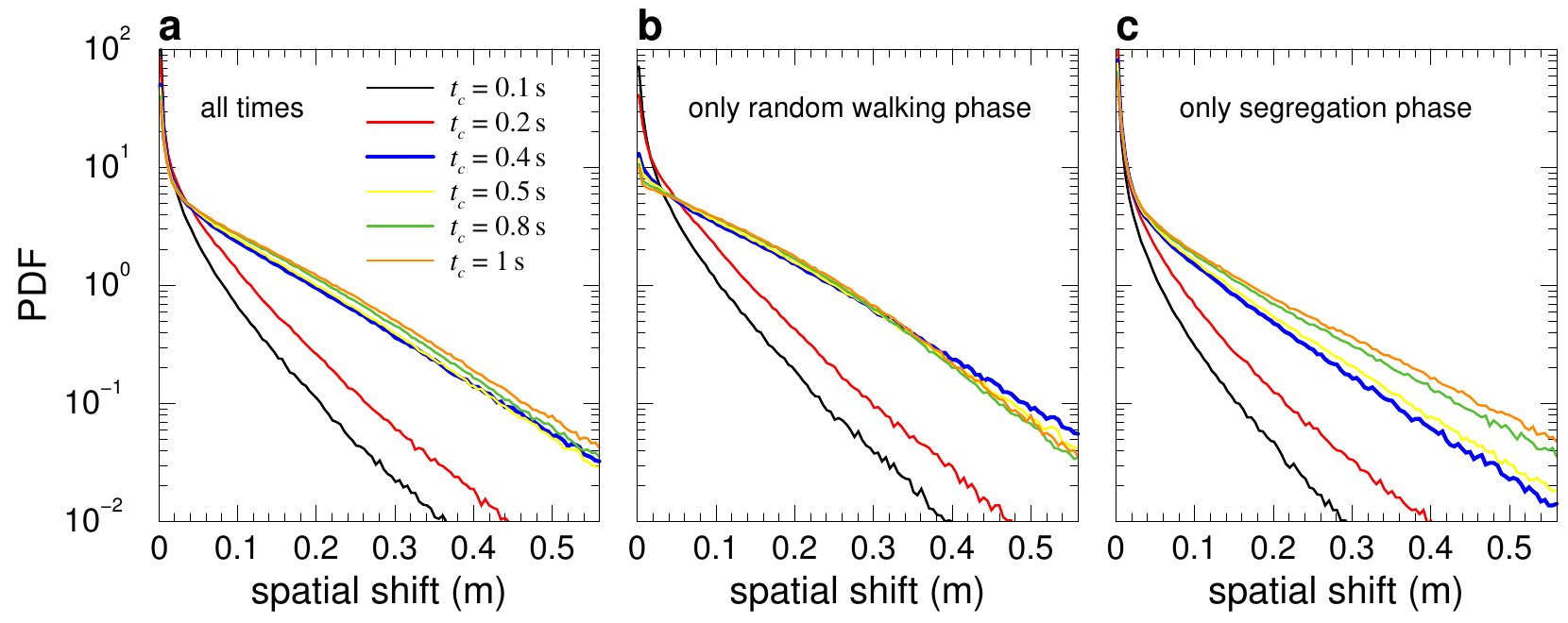}
\end{figure*}
\paragraph*{S4~Fig.}
{\textbf{Probability density function of the spatial shift} of the reconstructed positions of an
	individual at a given time for a value of $t_c > 0$, with respect to his/her position when
	no reconstruction is carried out ($t_c = 0$\,s), for all the individuals of all sessions, (a) at all times, (b) in the random walk phase only and (c) in the segregation phase only.
	For small values of $t_c$ the spatial shift is smaller in the segregation phase
	(around 10\,cm smaller) because the speed is generally smaller during this phase.
	This is not the case for $t_c = 0.8$ and 1\,s, where the homogenization of speed erases
	the difference between both phases. 	}
\label{decalage}
\FloatBarrier

\newpage

\subsection*{Supplementary tables}

\begin{table}[htp]
	 \begin{adjustwidth}{-3cm}{}
	\begin{center}
	{\small 
			\centering
			\begin{tabular}{c | lllllll | l}
				\hline \hline
				\Big. $k$ & \;1 & \;3 & \;5 & \;7 & \;9 & 11 & 13 & Total  \\
				\hline 
				\Big. Classic segregation & 27 (22) & 26 (20) & 27 (22) & 27 (21) & 25 (19) & 24 (19) & 28 (0)  & \, 184 (123)\\
				Segregation in 2 clusters & 17 & 17 & 17 & 17 & 17 & 17 & 17 & \, 119 \\
				\hline \hline
			\end{tabular}
		}
	\end{center}
\end{adjustwidth}
\end{table}
\vspace{-0.5cm}
\paragraph*{S1 Table.}	\label{Table1}
{The 303 segregation sessions, arranged by values of $k$.
	``Segregation in 2 clusters'' refers to the case where participants were explicitly asked to form two
	groups at the end of a session. Numbers between parentheses denote sessions carried
	out in September 2015. Other sessions were performed in June 2016.}

\vskip 1.5cm
%% Updated Table
\begin{table}[h!]
	%\rowcolors{2}{gray!25}{white}
	\centering
	\begin{tabular}{| c | c || c | c |}
		\hline
		Parameter & Value & Parameter & value  \\ \hline
		$v_0$ & 1.025 m/s & $\tau_0$ & 0.9 s \\ \hline
		$l_{\rm w}$ & 1.4 m & $l_{\rm h}$ & 1 m \\ \hline
		$r_{\rm w_c}$ & 2.5$\, l_{\rm w}$ & $r_{\rm h_c}$ & 1.6$\, l_{\rm h}$ \\ \hline
		$a_{\rm w}$ & 1.23 m/s$^2$ & $a_{\rm h}$ & 0.9 m/s$^2$ \\ \hline
		$a_{\rm w_0}$ & 0.865 & $a_{\rm w_1}$ & 0.694 \\ \hline
		$a_{\rm w_4}$ & -0.141 & $\sigma_{\rm 0}$ & 0.24 m/s$^{3/2}$ \\ \hline
	\end{tabular}
\end{table}
\vspace{-0.5cm}
\paragraph*{S2 Table.}	\label{Table2}
{Parameter values in the random walk phase; see Eq~(8-17).}

\vskip 1.5cm
%% Updated Table
\begin{table}[h!]
	%\rowcolors{2}{gray!25}{white}
	\centering
	\begin{tabular}{| c | c || c | c |}
		\hline
		Parameter & Value & Parameter & value  \\ \hline
		$v_{\rm beep}$ & 0.4 m/s & $\tau_{\rm beep}$ & 5 s \\ \hline
		%$v_{\rm noise}$ & 0.16 m/s & $\tau_{\rm noise}$ & 0.9 s \\ \hline
		$l_{\rm w}$ & 0.5 m & $l_{\rm h}$ & 0.2 m \\ \hline
		$r_{\rm w_c}$ & $l_{\rm w}$ & $r_{\rm h_c}$ & $l_{\rm h}$ \\ \hline
		$a_{\rm w}$ & 2.8 m/s$^2$ & $a_{\rm h}$ & 1.5 m/s$^2$ \\ \hline
		$a_{\rm w_0}$ & 0.865 & $a_{\rm w_1}$ & 0.694 \\ \hline
		$a_{\rm w_4}$ & -0.141 & $\sigma_{\rm 0}$ & 0.24 m/s$^{3/2}$ \\ \hline
		$\tau$ & 3 s & & \\ \hline
	\end{tabular}
\end{table}
\vspace{-0.5cm}
\paragraph*{S3 Table.}	\label{Table3}
{Parameter values in the segregation phase; see Eq~(8-19). }

\FloatBarrier\newpage
%%%%%%%%%%%%%%%%%%%%%%%%%%%%%%%%%%%%%%%%%%%%%%%%%%%%%%%%%%%%%%%%%%%%%%%%%%%%%%%%%%%%%%%%%

\subsection*{Supplementary videos}

\vskip 0.4cm
\paragraph*{S1 Video.} {\label{S1_video}
	An experimental run for $k=1$, starting with the initial random walk phase (duration of typically 45\,s; only 10\,s shown here) followed by the segregation phase, during which the beeping subjects have their corresponding colored circle showing  a white boundary. }

\paragraph*{S2 Video.} {\label{S2_video}
	An experimental run for $k=3$, starting with the initial random walk phase (duration of typically 45\,s; only 20\,s shown here) followed by the segregation phase, during which the beeping subjects have their corresponding colored circle showing  a white boundary.}

\paragraph*{S3 Video.} {\label{S3_video}
	Comparison between an experimental run (left) and a model simulation (right) for $k=3$. After the random walk period (15\,s shown here), the beeping subjects have their corresponding colored circle showing  a white boundary. The final 3-group structure is shown at the end of the segregation phase (see also  \nameref{final_config_1}).}

\paragraph*{S4 Video.} {\label{S4_video}
	An experimental run for $k=9$, in the case where subjects were additionally instructed to separate in two clearly identifiable groups, starting with the initial random walk phase (duration of typically 45\,s; only 8\,s shown here) followed by the segregation phase, during which the beeping subjects have their  colored circle showing  a white boundary. As exemplified by this video (see also the  \nameref{final_config_2}), the additional instruction to form two clearly identifiable groups almost systematically leads to a complete separation.}

\end{document}